\documentclass[
reprint, 
amsmath, amssymb, 
superscriptaddress,
groupedaddress,
floatfix, 
prb, 
showkeys, 
]{revtex4-1}

\usepackage{soul}
\usepackage{graphicx}
\usepackage{dcolumn}
\usepackage{bm}
\usepackage{txfonts}
\usepackage[usenames]{color}
\usepackage{appendix}

\usepackage{color}

\newcommand{\blue}[1]{{\color{blue} #1}}

\usepackage{tikz}
\usepackage[compat=1.1.0]{tikz-feynman}

\usepackage[colorlinks=true, citecolor=blue, linkcolor=blue, urlcolor=blue,
pdfpagelabels=true,
hyperindex=true, linktocpage, pdfa=true
]{hyperref}

\usepackage[autostyle]{csquotes}

\begin{document}



\title{Rescaling of Applied Oscillating Voltages in Small Josephson Junctions}

\author{Godwill Mbiti Kanyolo}
\thanks{gmkanyolo@gmail.com}
\altaffiliation{Department of Engineering Science, The University of Electro-Communications,
1-5-1 Chofugaoka, Chofu, Tokyo 182-8585, Japan.}

\author{Hiroshi Shimada\blue{$^*$}}



\begin{abstract}
The standard theory of dynamical Coulomb blockade [$P(E)$ theory] in ultra-small tunnel junctions has been formulated on the basis of phase-phase correlations by several authors. It was recently extended by several experimental and theoretical works to account for novel features such as electromagnetic environment-based renormalization effects. Despite this progress, aspects of the theory remain elusive especially in the case of linear arrays. Here, we apply path integral formalism to re-derive the Cooper-pair current and the BCS quasi-particle current in single small Josephson junctions and extend it to include long Josephson junction arrays as effective single junctions. We consider renormalization effects of applied oscillating voltages due to the impedance environment of a single junction as well as its implication to the array. As is the case in the single junction, we find that the amplitude of applied oscillating electromagnetic fields is renormalized by the same complex-valued weight $\Xi(\omega) = |\Xi(\omega)|\exp i\eta(\omega)$ that rescales the environmental impedance in the $P(E)$ function. This weight acts as a linear response function for applied oscillating electromagnetic fields driving the quantum circuit, leading to a mass gap in the thermal spectrum of the electromagnetic field. The mass gap can be modeled as a pair of exotic `particle' excitation with quantum statistics determined by the argument $\eta(\omega)$. In the case of the array, this pair corresponds to a bosonic charge soliton/anti-soliton pair injected into the array by the electromagnetic field. Possible application of these results is in dynamical Coulomb blockade experiments where long arrays are used as electromagnetic power detectors.

\end{abstract}

\pacs{Valid PACS appear here}
\maketitle


\section{\label{Intro}Introduction}

Since the pioneering theoretical work by Likharev et al., \cite{Likharev,Likharev2} $small$ Josephson junctions have been thought of as a dual system to $large$ Josephson junctions -- the roles of current and voltage are interchanged. In the case of $large$ Josephson junctions, their effective interaction with oscillating electromagnetic fields has been intensively studied, demonstrating their unique suitability for microwave-based applications such as the metrological standard for the \textit{Volt} (in terms of voltage Shapiro steps) and other microwave-based devices.\cite{Barone}
The dual system, on the other hand, holds enormous promise for complementary applications such as a metrological standard for the \textit{Ampere} in terms of the current Shapiro steps.\cite{Lisa2018}

However, observation of dual phenomena in $small$ junctions faces daunting experimental and theoretical challenges due, in part, to the lack of an approach that {\it consistently} covers both regimes.\cite{Likharev4}  
In particular, $small$ tunnel junctions are prone to quantum and thermal fluctuations -- their characteristics cannot be analyzed separate from their dissipative environment.\cite{Nyquist1928, FDT1951} As a consequence of Heisenberg uncertainty principle, their current-voltage $I-V$ characteristics is highly sensitive to energy changes in the environment.\cite{Shimada2016} Heuristically, tunneling of a single charge $e$ across a tunnel junction of capacitance $C$ and conductance $1/R$ is restricted unless the maximum energy it can absorb from zero-point fluctuations in the vacuum, $h/RC$, is sufficient to offset its own charging energy $e^2/2C = E_{\rm c}$ in the absence of other energy sources, where $h$ is Planck's constant. Thus, the heuristic condition for Coulomb blockade is $h/RC < E_{\rm c}$, which corresponds to $R_{\rm Q} < R$, where $R_{\rm Q} = h/e^2$ is the quantum resistance.

In the case of Josephson junctions, quantum tunneling is due to paired electrons (Cooper pairs). Thus, two distinct ratios parameterize the Josephson junction at zero temperature: 1) The phase/charge regime characterized by the ratio $E_{\rm J}/E_{\rm c}$ where $E_{\rm J}$ is the Josephson coupling energy and $E_{\rm c}$ the charging energy, which are simply the coefficients of the potential energy and kinetic energy respectively in the Hamiltonian of the Josephson junction; and 2) the superconducting/Coulomb blockade regime characterized by the aforementioned ratio $R_{\rm Q}/R$ where $R$ is the real part of the environmental impedance, $Z(\omega)$. In this picture, the $small$ junction is defined in the regime $E_{\rm J}/E_{\rm c} < 1$ and simultaneously $R_{\rm Q}/R < \kappa^2$, where $\kappa = 2$ is the number of electrons in a single Cooper pair.

Consequently, lifting of Coulomb blockade occurs when other sources of energy are present. For instance, energy is easily supplied by thermal fluctuations $\beta^{-1} = k_{\rm B}T > 0$, strong coupling between pairs across the junction $E_{\rm J} \gg  E_{\rm c}$ or external constant voltages $V_{\rm x} > V_{\rm cb} \sim  E_{\rm c}$ above the Coulomb blockade threshold voltage, thus resulting in current-voltage characteristics highly dependent on these environmental parameters. Formally, this implies that the action for large junctions $iS(\phi) \simeq \ln\left [\prod_n \int D \phi_{\rm n} \exp i\sum_{n} S_n(\phi, \phi_n) \right] $ is effective, meaning it emerges from tracing out environmental degrees of freedom $\sum_n S_n(\phi, \phi_n)$ that act as energy sources for the tunnel junction. This leads to a theory of dynamical Coulomb blockade [$P(E)$ theory] in single small Josephson junctions formulated on the basis of $ \left \langle \phi(t) \phi(0) \right \rangle = \mathcal{Z}^{-1} \int D\phi \left [ \phi(t) \phi(0) \right ]\exp iS(\phi)$ phase correlations\cite{girvin1990, devoret1990, Falci1991, Ingold_Nazarov1992} where tunneling across the barrier is influenced by a high impedance environment treated within the Caldeira-Leggett model.\cite{Caldeira-Leggett1983} 

$P(E)$ theory has successfully been tested to a great degree of accuracy in a myriad of experiments.\cite{pierre2001, holst1994, cleland1992, geerligs1989, delsing1989} This has lead to its widespread application in describing progressively complex tunneling processes such as dynamical Coulomb blockade in small Josephson junctions and quantum dots.\cite{Devoret2001,Qdots2001} Moreover, owing to significant improvement in microwave precision measurement technology such as near-quantum-limited amplification\cite{Par_Amp2007, Joyez_Group2018} 
and progress in theory, recently published works suggest novel features in the $P(E)$ framework ranging from time reversal symmetry violation\cite{Safi2011} and Tomonaga-Luttinger Liquid (TLL) physics,\cite{TLphysics2013} to renormalization of electromagnetic quantities appearing in the $P(E)$ function.\cite{Grabert2015, PowerRen2015, E_Jren2015, E_Jren2019} Despite this progress, aspects of the theory remain elusive especially in the case of linear arrays.

Here, we apply path integral formalism to derive the Cooper-pair current and the BCS quasi-particle current in single small Josephson junctions. We consider renormalization effects of applied oscillating voltages due to wavefunction renormalization/Lehmann weights\cite{Lehmann1954} that rescale the environmental impedance of the single junction as well as the array.
The array is treated as infinitely long\cite{Bakhvalov1989} and transformed into an effective circuit. As is the case for the single junction, we show that the Lehmann weight, $\Xi(\omega) = \exp(-\Lambda^{-1})\exp [-\beta M(\omega)]\exp i\beta \varepsilon_{m}$ also acts as a linear response function for oscillating electromagnetic fields, and can be interpreted as the probability amplitude of exciting a `particle' of mass $M$ from the junction ground state by the radio-frequency (RF) field.\cite{Kanyolo_Berry} The quantum statistics of this `particle' are determined by the argument $\beta \varepsilon_{m}$ where $\varepsilon_{m}$ is identified as the Matsubara frequency.\cite{Matsubara1955} In the case of the infinite array, this `particle' corresponds to a bosonic charge soliton injected into the array. Possible application of these results is in accurately determining the absorbed RF power in dynamical Coulomb blockade experiments especially where long arrays are used as on-site electromagnetic power detectors.\cite{Liou2014, LT28}

The paper is organized as follows:

Section \ref{Renormalization} deals with the rescaling of oscillating voltages applied on single junctions. In the subsections, \ref{Renorm_intro} explains the basis for this rescaling, \ref{Impedance renormalization} introduces the finite temperature propagator and the environmental impedance as Green's function, \ref{Vacuum Excitation} interprets the impedance Lehmann weight as a complex-valued probability amplitude for applied oscillating voltages (RF field) exciting a `particle' with quantum statistics given by the argument of the factor, or equivalently as a linear response function and the RF field as the external force leading the full expression for the current-voltage characteristics that includes the RF field and renormalization effects. 

Finally, section \ref{Sec: Arrays} considers the Lehmann weight in infinitely long arrays.
The difference from the single junction is an additional Lehmann weight $\exp (-\Lambda^{-1})$ representing a finite range of the electromagnetic field due to the presence of a charge soliton of length $\Lambda$ injected into the infinitely long array.  

Note that, units where Planck's constant, Swihart velocity\cite{Swihart1961} and Boltzman constant are set to unity ($\hbar = \bar{c} = k_{\rm B} = 1$) and Einstein summation convention are used through out unless otherwise stated with ${\rm diag}\left \{ \eta_{\mu\nu} \right \} = (1, -1, -1, -1)$ the Minkowski space-time metric and $\eta_{\sigma\mu}\eta^{\sigma\nu} = \delta_{\mu}^{\nu}$ the Kronecker delta symbol.

\section{\label{Renormalization} Rescaling of Oscillating Voltages Applied on Single Junctions}

\subsection{\label{Renorm_intro} Introduction}

\begin{figure}[b!]
\begin{center}
\includegraphics[width=0.6\columnwidth,clip=true]{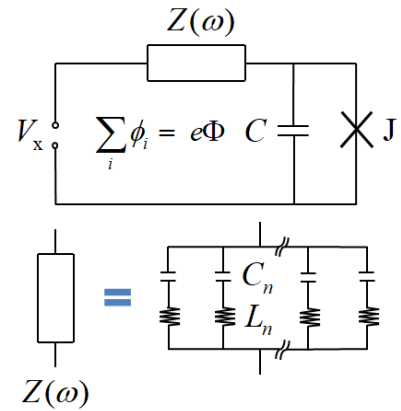}
 \caption{Mesoscopic tunnel junction, J with capacitance $C$ driven by a voltage source $V_{\rm x}$ via an environmental impedance $Z(\omega)$ composed of infinite number of parallel $L_n$ $C_n$ circuits. The circuit stores a flux $e\Phi = \sum_i \phi_i = \phi_{\rm J} + \phi_{\rm x} + \phi_{\rm z}$ related to a topological potential $A(t)$ by $\int_{-\infty}^{t}\,ds A(s) = \Phi(t)$.}
\label{Caldeira_fig}
\end{center}
\end{figure}

Within the Caldeira-Leggett model\cite{Caldeira-Leggett1983}, the environment of a dissipative voltage-biased single junction shown in Fig. \ref{Caldeira_fig} is modeled by the action $S_{\rm z} = \int dt\, \mathcal{L}_{\rm z}$, where the Lagrangian is given by,
\begin{subequations}
\begin{align}
\mathcal{L}_{\rm z} = \frac{C}{2e^2}\left ( \frac{\partial\phi'}{\partial t} \right )^2 + \sum_{n = 1}^{k} \left \{ \frac{C_{n}}{2e^2}\left (\frac{\partial\phi_{n}}{\partial t} \right )^2 - \frac{L_{n}^{-1}}{2e^2}\left (\phi_{n} - \phi'\right )^2 \right \}, 
\label{l_CL_eq}
\end{align}
where $e$ is the electric charge, $C = \varepsilon_{0}\varepsilon_{\rm r}\mathcal{A}/d_{\rm eff}$ is the junction capacitance, $\phi' \equiv \phi_{\rm J} - \phi_{\rm x} - e\Phi$ where $\phi_{\rm J}$, $\phi_{\rm x}$, $\phi_{\rm z}$ and $e\Phi$ are the phases associated with the the voltage drop at the junction $V_{\rm J}$, voltage source (external voltage) $V_{\rm x}$ and the flux stored by the circuit respectively, $\phi_{n}$ is the bath degrees of freedom represented by $k$ coupled (via $\phi$) $L_{n}C_{n}$ oscillators constituting the environment. Note that $\phi'$ is the phase defined for convenience, to shorten the expression of the Lagrangian, avoiding always writing the full expression $\phi_{\rm J} - \phi_{\rm x} - e\Phi = \phi - e\Phi$.

The effective action,
\begin{multline}
S'_{\rm z} = -i\ln \int \prod_{n =1}^{k} D\phi_n\exp iS_{\rm z}(\phi_{n}, \phi') \\
= \int dt \left \{ \frac{C}{2e^2}\left ( \frac{\partial\phi'}{\partial t} \right )^2 - \frac{1}{4\pi e^2}\int ds\,\phi'(s)\left [ \frac{\partial Z^{-1}(t-s)}{\partial t} \right ]\phi'(t)  \right \}\\
= \frac{2\pi}{2e^2} \int d\omega\, \phi'(\omega)i\omega Z^{-1}_{\rm eff}(\omega)\phi'(-\omega)
\label{S_eff_CL_eq}
\end{multline}
\end{subequations}
requires the impedance Green's function in the $P(E)$ function to be modified by a wavefunction renormalization (Lehmann) weight,
\cite{Lehmann1954}),
\begin{subequations}\label{P(E)_ren_eq}
\begin{align}
P_{\kappa}(E) = \frac{1}{2\pi}\int dt \exp\left (\kappa^2\mathcal{J}(t) + iEt \right )\\
\mathcal{J}(t) = \frac{2e^2}{2\pi}\int \frac{d\omega}{\omega} {\rm Re}\left \{Z_{\rm eff}(\omega)\right \}\frac{\exp(-i\omega t) - 1}{1 - \exp(-\beta \omega)}\\
Z_{\rm eff}(\omega) = [Z^{-1}(\omega) + i\omega C]^{-1} = \Xi(\omega)Z(\omega)
\end{align}
\end{subequations}
where $\beta$ is the inverse temperature, $\kappa e = 1e, 2e$ is the quasi-particle, Cooper-pair charge, $E$ is the energy exchanged between the junction, $\Xi(\omega)$ is a Lehmann weight and $L_{n}C_{n}$ circuits acting as the environment, $\omega$ is the Fourier transform frequency that also plays the role of the thermal photon frequency at finite temperature. 

It is known -- at least since the work of Callen and Welton\cite{FDT1951} -- that the (causal) response function\footnote{Appendix \ref{App: linear_response}} $\Xi(\omega) \equiv \int_{-\infty}^{+\infty} dt\, \theta(t)\chi(t)\exp(i\omega t) dt\, $ for a system driven by oscillating electromagnetic fields appears as the coefficient\footnote{This co-efficient can be computed using the driven system's equation of motion.} of the black body spectrum. Consequently, this requires that the response $V'_{\rm RF}(t)$ as seen by the junction $\rm J$ in Fig. \ref{Caldeira_fig} be a weighted function of  $\chi(t)$ and the applied oscillating voltage $V_{\rm RF}(t): V'_{\rm RF}(t) = \int_{-\infty}^{t} ds\, \chi(t-s)V_{\rm RF}(s)$. Therefore, to accurately describe the $I$--$V$ characteristics of ${\rm J}$ driven by an applied oscillating voltage $V_{\rm RF}(t)$, it is not enough to simply rescale the impedance $Z(\omega)$ in the $P(E)$ function: the amplitude and phase of the applied oscillating voltage $V_{\rm RF}$ has to be renormalized accordingly. 

In subsequent sections, we first consider tracing our steps from standard quantum electrodynamics (QED) and ease our way into circuit-QED and hence $P(E)$ theory. We then proceed to introduce the finite temperature propagator for the junction and consider how the Lehmann weight arises for the impedance in $P(E)$ theory, and its implications for single junctions and long arrays driven by $V_{\rm RF}(t)$. We find that, a finite time varying flux $\Phi(t)$ stored by the circuit consistently implements the aforementioned wavefunction renormalization by guaranteeing the circuit responds linearly to $V_{\rm RF}(t)$. 

\subsection{Connection of \textit{P(E)} theory to quantum electrodynamics (QED).}

Here, we shall connect the Caldeira-Leggett model to QED. We shall find out that circuit-QED is merely the 1 dimensional space time version of QED. This also allows us to link the propagator introduced in eq. (\ref{propagator_eq}) with the photon propagator in QED.  

The Fourier transform of the summed terms in Caldeira-Leggett action given in eq. (\ref{l_CL_eq}) is $S_{0} + S_{\rm int}$ where, 
\begin{multline}
    S_{0} = \frac{2\pi}{2e^2 }\sum_n\int d\omega C_n\phi_n(\omega)\left [\omega^2 - \omega_n^2 \right ]\phi_n(-\omega)\\ 
    + \frac{2\pi}{2e^2}\sum_n \frac{1}{L_{n}}\int d\omega \phi_n(\omega)\phi'(-\omega) + O(\phi'^2),
    \label{eq: CL}
\end{multline}
$O(\phi'^2)$ is a term with $\phi'^2$ that we initially neglect, $\phi_n$ are the Caldeira-Leggett phases of $L_nC_n$ circuits in Fig. \ref{Caldeira_fig}, $\omega_n = 1/L_nC_n$ and the interaction term $S_{\rm int}$ is given by,
\begin{multline}
S_{\rm int} = \frac{2\pi}{2e^2 }\int d\omega \omega^2 C\phi'(\omega)\phi'(-\omega)\\
+ \frac{2\pi}{e}\int d\omega I_{\rm F}(-\omega)\phi'(\omega),
\end{multline}
where $I_{\rm F}$ is the fluctuation current which we shall later set, $I_{\rm F} = 0$.

Integrating out the fluctuating degrees of freedom, $\phi_n$, 
\begin{subequations}
\begin{align}
     \prod_n \int D\phi_n \exp(iS_{0}) = \exp(iS_{0}'),
     \label{Int_out_eq}
\end{align}
we arrive at,
\begin{align}
    S_{0}' = \frac{(2\pi)^2}{4\pi}\int d\omega \phi'(-\omega) G^{-1}(\omega)\phi'(\omega),\\
    G^{-1}(\omega) = \frac{1}{e^2}\sum_n \frac{1}{L_n}\frac{\omega_n^2}{\omega^2 - \omega_n^2} \equiv i\omega e^{-2}Z^{-1}(\omega),
    \end{align}
\end{subequations}
where we have converted the Green's function $G(\omega)$, from the $\phi_n$ degrees of freedom, to the environmental impedance $Z(\omega)$. 

Proceeding to combine the two actions yields,
\begin{multline}
S_{0}' + S_{\rm int}
= \frac{2\pi}{2e^2}\int d\omega \phi'(\omega)\left [e^2G^{-1}(\omega) + \omega^2C \right ]\phi'(-\omega)\\
+ \frac{2\pi}{e}\int d\omega I_{\rm F}(-\omega)\phi'(\omega).
\label{effective_eq}
\end{multline}
By defining the electromagnetic vector potential $A_{\mu} = (V, \vec{A})$, the electric field $\vec{E} = \partial \vec{A}/\partial t - \vec{\nabla}V$, the magnetic field $\vec{B} = \vec{\nabla}\times\vec{A}$ and the fluctuation current density $J^{\mu}_{\rm F}$, it can be seen that the interaction term given by $S_{\rm int}$ above is actually Maxwell’s action in disguise,
\begin{multline}
    S_{\rm int} \propto  \int dt d\mathcal{A} dl \left [ \frac{\varepsilon_0\varepsilon_{\rm r}}{2}(\vec{E}\cdot\vec{E} - \vec{B}\cdot\vec{B}) + eJ_{\rm F}^{\mu}A_{\mu} \right ]\\
    = \frac{1}{2}\int \frac{d^{\,4}k}{(2\pi)^4} \varepsilon_0\varepsilon_{\rm r} A^{\mu}(k)G_{\mu\nu}^{-1}A_{\nu}(-k)\\
    + e\int \frac{d^{\,4}k}{(2\pi)^4} J^{\mu}_{\rm F}(-k)A_{\mu}(k),
\end{multline}
with the conditions $\vec{B} = 0$, $\partial \phi'/\partial t = -el\vec{n}\cdot\vec{E}$, $\phi' = e\int dl\,\vec{n}\cdot\vec{A}$, $C = \varepsilon \mathcal{A}/l$ and $\int d\mathcal{A}\vec{n}\cdot\vec{J}_{\rm F} = I_{\rm F}$ where $\vec{n} = (1, 0, 0)$ is the normal vector to the junction barrier, $l \equiv d_{\rm eff}$ the effective barrier thickness and $\mathcal{A}$ the junction area. The last term is the Fourier transform of the action where $k \equiv k^{\mu} = (\omega, \vec{k})$ is the photon energy-momentum satisfying $k^2 \equiv k^{\mu}k_{\mu} = \varepsilon^2$ and the Green's function in 1 + 3 space-time takes the form,\cite{Zee2010}
\begin{align}
    G_{\mu\nu} = \lim_{\varepsilon \rightarrow 0} \frac{-\eta_{\mu\nu} + k_{\mu}k_{\nu}/\varepsilon^2}{k^2 - \varepsilon^2}.
\end{align}

Integrating out $\phi'(\omega)$ degrees of freedom in $S_{0} + S_{\rm int}$ given by eq. (\ref{effective_eq}) as in eq. (\ref{Int_out_eq}) leads to a circuit-QED term,
\begin{align}
    \frac{\pi}{e^2}\int d\omega I_{\rm F}(\omega)G_{\rm eff}(\omega)I_{\rm F}(-\omega)
\end{align}
where $G_{\rm eff}^{-1} = G_{\omega}^{-1} + e^{-2}\omega^2C$ is reminiscent of the Coulomb interaction term in QED,
\begin{align}
    \alpha \int  \frac{d^{\,4}k}{(2\pi)^3} J^{\mu}_{\rm F}(k)G_{\mu\nu}(k)J^{\nu}_{\rm F}(-k),
\label{Coulomb_D}
\end{align}
where $\alpha = e^2/4\pi \varepsilon_0\varepsilon_{\rm r}$ is the fine structure constant. The QED term is obtained in a similar fashion by integrating out $A_{\mu}$ instead of $\phi'$. Nonetheless, both expressions are essentially describing the same process. The difference is the dimensionality of the theory: QED is in 1 + 2 space-time dimensions whereas circuit-QED is solely in the time dimension (circuit-QED). Thus, we have showed that $1/\omega^2C$ and hence $G(\omega)$ both have an interpretation as photon propagator in circuit-QED.

However, a question still remains: is there any significance of this trivial Fourier space transformation given by $G^{-1}(\omega) \rightarrow G_{\rm eff}^{-1}(\omega)$? We notice that we can define a factor $\Xi(\omega) = G(\omega)/G_{\rm eff}$ and claim that this factor renormalizes the propagator $G(\omega)$ (of the $\phi_n$ degrees of freedom) to $G_{\rm eff}(\omega)$ due to the presence of the Maxwell term, $S_{\rm int}$. Since this renormalization takes a photon propagator into a different photon propagator, the Feynman rules to calculate it resemble photon self-energy interactions. 

In particular, in self-energy interactions, the photon polarizes the QED vacuum by creating electron-positron pairs which subsequently annihilate. Such pairs can be created an infinite number of times, thus the contribution to the amplitude of all the processes takes the form: $G_{\rm eff} = G + GUG + GUGUG \cdots = G/(1-UG) = 1/(G^{-1} - U)$, where $G$ is the photon propagator and $U$ the vacuum polarization energy (interaction term).\cite{Zee2010} Bearing this in mind, we formulate the following Feynman rules for the propagator:
\begin{enumerate}

    \item  The photon propagator $G(\omega) = e^2Z(\omega)/i\omega$ is represented by:
        \feynmandiagram [small, horizontal=a to b] {a -- [photon] b};
    ;
      
    \item The vacuum polarization energy term, $U(\omega) = e^{-2}\omega^2C$ is represented by:
        \feynmandiagram [small, horizontal=a to d] {
        b [blob],};
    ;
    
    \item Therefore, the leading order interaction term,\\ $G(\omega)U(\omega)G(\omega)$ is drawn as,
     \feynmandiagram [small, horizontal=a to d] {
    a -- [photon] b [blob] c -- [photon] d,
    };, 
\end{enumerate}
where time increases from left to right. Note that diagram 1 reads as follows: A photon of energy $\omega$ is created, propagates with a probability amplitude $G(\omega)$ and annihilates at a later time. Thus the amplitude must be assigned a photon creation operator and an annihilation operator, $a^{\dagger}$ and $a$ respectively and, in mathematical form, it should be written as $aG(\omega)a^{\dagger}$.

Likewise, diagram 3 represents an interaction whereby a photon of frequency $\omega$ is produced by acting on the vacuum state with $a^{\dagger}$, it propagates with an amplitude given by $G(\omega)U(\omega)G(\omega)$ then it annihilates by acting on the vacuum with $a(\omega)$. We emphasize that reversing $\omega$ reverses the aforementioned processes. This implies that the operators themselves should also be defined accordingly as,
\begin{subequations}
\begin{align}
a(-|\omega|) = a^{\dagger},\\ 
a(|\omega|) = a,\\
[a(\omega), a(\omega')] =\left [ \theta(\omega')-\theta(\omega) \right ]\delta_{\omega, -\omega'},
\end{align}
\end{subequations}
where $\theta(\omega)$ is the Heaviside function. This clearly displays the roles of the positive and negative frequencies. Note that negative frequencies are allowed since we are interested in energy differences due to single photon emission and absorption processes. For instance, processes where a photon is created before annihilation are related to processes where a photon is annihilated before creation by reversing the sign of the frequency $\omega$ and re-ordering the $a^{\dagger}, a$ operators appropriately.\footnote{This procedure is invalid when arguments for time-reversal asymmetry e.g. discussed in ref. \citenum{Safi2011} apply.}

\subsection{\label{Impedance renormalization} Connection to \textit{P(E)} theory: The finite Temperature Green's Function and Propagator} 

Observe that a straightforward regularization procedure verifies $Z_{\rm eff}(\omega)$ plays the role of effective Green's function of the $P(E)$ function,
\begin{multline}\label{pert_expand_eq}
Z_{\rm eff}(\omega) = Z(\omega) + Z(\omega)[-i\omega C Z(\omega)] + Z(\omega)[-i\omega C Z(\omega)]^2 \\
+ \cdots
+ Z(\omega)[-i\omega C Z(\omega)]^{n \rightarrow +\infty} = Z(\omega)\sum_{n = 0}^{+\infty}[-i\omega C Z(\omega)]^{n}\\
= \frac{Z(\omega)}{1 + i\omega C Z(\omega)} = \Xi(\omega)Z(\omega), 
\end{multline}
analogous to the renormalization of the propagator in QED which often leads to a (Lehmann) factor\cite{Lehmann1954, WFR_1951} analogous to $\Xi(\omega)$. 

To elucidate this, consider the finite temperature propagator for $S'_{0}$, given by
(\feynmandiagram [large, horizontal=a to b] {a -- [photon] b};), 
\begin{widetext}
\begin{align}\label{D_eq}
D^{\kappa}(\omega) = \frac{-\kappa^2}{2\pi i}\left \{  G(\omega)\left \langle a(\omega)a(-\omega) \right \rangle + G(-\omega)\left \langle a(-\omega)a(\omega) \right \rangle \right\}
\end{align}
\end{widetext}
where $\kappa e = 1e, 2e$ is the Cooper pair, BCS quasi-particle charge. The prefactor $\kappa^2/2\pi i$ included for ease of comparison later with the $P(E)$ theory. The processes with $D^{\kappa *}(\omega) = \frac{-\kappa^2}{2\pi i}\left \{  G(-\omega)\left \langle a(\omega)a(-\omega) \right \rangle \right \}$ $+ \frac{-\kappa^2}{2\pi i}\left \{G(\omega)\left \langle a(\omega)a(-\omega)\right \rangle \right \}$ are forbidden since represent {\it actual} negative frequency photons.

We proceed to include the effect of $S_{\rm int}$ from a finite tunnel junction impedance by introducing the potential $U(\omega) = e^{-2}(\omega^2 C - \sum_n L_n^{-1}) = -e^{-2}i\omega y(\omega)$ where we have restored the $O(\phi'^2)$ term from eq. (\ref{eq: CL}). The effective propagator is given by the sum of all the single photon interactions at the junction due to $U(\omega)$ and is proportional to 
$\left \langle a(\omega)a(-\omega) \right \rangle$ for positive and $\left \langle a(-\omega)a(\omega) \right \rangle$ for negative frequencies [eq.  (\ref{D_eq})], we find
\begin{subequations}
\begin{widetext}
\begin{multline}\label{D_eff_eq}
D_{\rm eff}^{\kappa}(\omega) = \frac{-\kappa^2}{2\pi i}\left \{ G(\omega) + G(\omega)U(\omega)G(\omega) + ... \right \}\left \langle a(\omega)a(-\omega) \right \rangle\\
+ \frac{-\kappa^2}{2\pi i}\left \{ G(-\omega) + G(-\omega)U(-\omega)G(-\omega) + ... \right \}\left \langle a(-\omega)a(\omega) \right \rangle
= \frac{-\kappa^2}{2\pi i}G(\omega)[1-U(\omega)G(\omega)]^{-1} \left \langle a(\omega)a(-\omega) \right \rangle\\
+ \frac{-\kappa^2}{2\pi i}G(-\omega)[1-U(-\omega)G(-\omega)]^{-1} \left \langle a^{\dagger}a\right \rangle_{-\omega}
= \frac{-\kappa^2}{2\pi i}\left \{G_{\rm eff}(\omega)\left \langle a(\omega)a(-\omega) \right \rangle
+ G_{\rm eff}(-\omega) \left \langle a(-\omega)a(\omega) \right \rangle\right \}.
\end{multline}
\end{widetext}
The term proportional to $[U(\pm \omega)G(\pm \omega)]^n$ is the finite temperature propagator for the photon interacting $n$ times with the junction impedance and the perturbation series 
\begin{multline}\label{amplitude_eq}
\sum_{n = 0}^{+\infty}[U(\pm \omega)G(\pm \omega)]^n = [1 + y(\pm \omega) Z(\pm \omega)]^{-1}\\
= 1 - y(\pm \omega) Z_{\rm eff}(\pm \omega) \equiv \Xi(\pm \omega),
\end{multline}
\end{subequations}
is computed by analytic continuation of the series $1 + x + x^2 \cdots + x^n \rightarrow [1-x]^{-1}$, where $x$ is given by the diagram, 
\feynmandiagram [small, horizontal= b to d] {
   b [blob] c -- [photon] d,
    };.
Thus, the effective Green's function $G_{\rm eff}(\omega)$ becomes,	
\begin{multline*}
\feynmandiagram [small, horizontal=a to b] {a -- [photon] b}; + \feynmandiagram [small, horizontal=a to d] {a -- [photon] b [blob] c -- [photon] d,}; + \feynmandiagram [small, horizontal=a to f] {a -- [photon] b [blob] c -- [photon] d [blob] e -- [photon] f,}; + \cdots\\
= \feynmandiagram [small, horizontal= c to d] {c -- [photon] d,};\times\frac{1}{\left (1 - \feynmandiagram [small, horizontal= b to d] {b [blob] c -- [photon] d,}; \right )}
= \frac{1}{\left ([\feynmandiagram [small, horizontal= c to d] {c -- [photon] d,};]^{-1} - \feynmandiagram [small, horizontal=a to d] {b [blob],}; \right )}.
\end{multline*} 
Comparing eq. (\ref{D_eq}) to (\ref{D_eff_eq}), we find that $G(\omega)$ is rescaled to $G_{\rm eff}(\omega) = [G^{-1}(\omega)-U(\omega)]^{-1}$. Consequently, the effective action is given by
\begin{multline}\label{S_eff_eq}
\left.\ S'_{\rm z}\right\vert_{I_{\rm F}=0} = \left.\ S_{0}' + S_{\rm int} \right\vert_{I_{\rm F}=0}\\ =\pi\int_{-\infty}^{+\infty} d\omega\, \phi'(\omega)G^{-1}_{\rm eff}(\omega)\phi'(-\omega),
\end{multline}
which is equivalent to eq. (\ref{S_circuit_eq}) with $I_{\rm F} = 0$.

Indeed we recover $\mathcal{J}(t) = e^2\left [ D^{\kappa}_{\rm eff}(t) - D^{\kappa}_{\rm eff}(0) \right ]$, where $D^{\kappa}_{\rm eff}(t) = \int d\omega\, D^{\kappa}_{\rm eff}(\omega)\exp(-i\omega t)$ is the Fourier transform of $D^{\kappa}_{\rm eff}(\omega)$,\footnote{Also confer eq.  (\ref{S_circuit_eq}) and (\ref{propagator_finite_eq}).} by substituting $\left \langle a(-\omega)a(\omega) \right \rangle = n(-\omega) = [\exp (-\beta \omega)-1]^{-1}$ and $\left \langle a(\omega)a(-\omega) \right \rangle = n(\omega) + 1 = [1-\exp (-\beta \omega)]^{-1}$ in eq.  (\ref{D_eq}) and (\ref{S_eff_eq}), where the analytic continuation, 
\begin{equation}\label{regularize_eq}
\sum_{n = 0}^{+\infty} \exp (\mp \beta n |\omega|) = \frac{1}{1 - \exp(\mp \beta |\omega|)}
\end{equation}
has been used to regularize the divergent sum when calculating averages, $\left \langle \cdots \right \rangle$ for negative frequencies.\\

\subsection{\label{Vacuum Excitation} Vacuum Excitation and Photon amplitudes}

Notice that in eq. (\ref{D_eff_eq}), the factor $[1-U(\omega)G(\omega)]^{-1} = \Xi(\omega) \equiv \left | \Xi(\omega) \right |\exp\left \{i\eta(\omega)\right \} = \left | \Xi(\omega) \right |\exp (i\beta\varepsilon_{m})$ either rescales $G(\omega)$ or the photon number thermal averages $\left \langle \cdots \right \rangle$, implying that the photon number states are modified by the junction impedance. Rearranging, we find
\begin{widetext}
\begin{multline}
G_{\rm eff}(\omega) \left \langle a(-\omega)a(\omega)\right \rangle_{\omega}
= G(\omega)\Xi(\omega)\left \langle a(-\omega)a(\omega)\right \rangle_{\omega}
= G(\omega)\left \langle \left | \Xi(\omega) \right |\exp\left \{i\eta(\omega)\right \}a(-\omega)a(\omega) \right \rangle_{\omega} \\
= G(\omega)\mathcal{Z}^{-1}_{\rm ph}\sum_{n = 0}^{\infty}\langle n|a(-\omega)a(\omega)
\exp\left \{-\beta \omega \left [ a(-\omega)a(\omega) + \frac{{\rm sgn}(\omega)}{2} \right ] + \ln\left [ \Xi(\omega) \right ]\right \}|n \rangle_{\omega}\\
\equiv G(\omega)\mathcal{Z}^{-1}_{\rm ph}
\sum_{n = 0}^{\infty}\langle n|a(-\omega)a(\omega)
\exp \left \{-\beta\left [ \omega [a(-\omega)a(\omega) + \frac{{\rm sgn}(\omega)}{2}] + M - i\varepsilon_{m} \right ]\right \}|n \rangle_{\omega},
\label{free_energy_eq}
\end{multline}
\end{widetext}
where ${\rm sgn}(\omega)$ is the sign function, $M -i\varepsilon_m = -\beta^{-1}\ln\Xi(\omega)$ is a gap in the electromagnetic energy spectrum. Taking $M(\omega)$ to be positive definite leads to $|\Xi(\omega)|^2\leq  1$. 

Thus, $M$ is the energy of a `particle' excited from the vacuum by the electromagnetic field where the probability that the vacuum will be excited is given by the Boltzmann distribution $|\Xi(\omega)|^2 = \exp\left \{ -\beta 2M \right \}$. Since this excitation carries electromagnetic energy, $|\Xi(\omega)|^2$ gives the fraction of electromagnetic power absorbed via excitations at the capacitor $C$ by the junction.\cite{Kanyolo_Berry}
We note that the complex nature of $\beta^{-1}\ln\Xi(\omega)$ is not a concern since we are already accustomed to shifting our frequencies or energies by an infinitesimal [e.g. $\omega$ to $\omega + i\varepsilon$ in eq. (\ref{Gen_Admittance_eq})].
Taking in eq. (\ref{amplitude_eq}) the impedance $Z(\omega) = R$ to be real and $y(\omega) = i\omega C$, corresponding to eq. (\ref{RCSJ_eq}), yields 
\begin{align}
\arctan{\omega RC} = \eta(\omega) = \beta\varepsilon_{m}.
\label{eta_eq}
\end{align}

We introduce quantum statistics of the `particle' by identifying $\varepsilon_{m} = (2m + 1)\pi\beta^{-1}$ or $\varepsilon_{m} = 2\pi m \beta^{-1}$ as the fermionic or bosonic Matsubara frequency\cite{Matsubara1955} respectively where $m \in \mathbb{Z}$ is an integer. This requires the oscillation period $2\pi/\omega$ of the electromagnetic field to greatly exceed the relaxation time $RC$ of the circuit, $2\pi/\omega \gg 2\pi RC$. Thus, when this condition is not satisfied, it leaves the possibility for `anyons'\cite{Chern_Simons1982} with exotic statistics. Consequently, we can take $\Xi(\omega)$ as the amplitude\footnote{Photon amplitude here refers to a wavefunction renormalization (Lehmann) weight, where the photon wavefunction is taken to be the $x$ component of the Riemann -- Silberstein vector\cite{Riemann_Silberstein}} that a photon of frequency $\omega$ is absorbed by the junction creating a `particle' of mass $M$ and statistics according to the Matsubara frequency $\varepsilon_{m}$. This realization, together with the fact that the field theory introduced in Appendix \ref{Josephson-Maxwell} lives in 1 + 2 dimensions, suggests that anyonic excitations cannot be ignored.\cite{Chern_Simons1982} 

Finally, notice that the exponent can be re-written conveniently as,
\begin{subequations}
\begin{multline}
\mathcal{G}^{-1}(M, i\varepsilon_m) + \frac{1}{2}\coth(\beta\omega/2) \\
= \mathcal{G}^{-1}(M, i\varepsilon_m) + \sum_{m = -\infty}^{+\infty} \frac{1}{2\pi mi - \beta \omega} \label{exponent_eq}
\end{multline}
where $\mathcal{G}(M, i\varepsilon_m) = \beta (M - i\varepsilon_m)$ is the Green's function of the `particle'. Thus, its excitation statistics can be computed as, 
\begin{align}
\sum_{m = -\infty}^{+\infty} \mathcal{G}(M, i\varepsilon_m) = \sum_{m = -\infty}^{+\infty}  \frac{\beta^{-1}}{M - i\varepsilon_m} 
\label{excitation_distribution_eq}
\end{align}
\end{subequations}
where eq. (\ref{excitation_distribution_eq}) is the inverse thermal Green's function of the `particle' with mass $M$. Whether these `particles' in the single junction are anything more than a tool to implement the renormalization scheme above remains to be investigated.

\subsubsection{\label{Sec: RF} Applied Alternating Voltages} 

In section \ref{Vacuum Excitation}, we have established that the fraction of photons absorbed by a tunnel junction is $|\Xi(\omega)|^2 = [1+y(\omega)Z(\omega)]^{-2}$. A straightforward way to experimentally measure $|\Xi(\omega)|^2$ is applying an external oscillating electric field in the form of ac voltage $V_{\rm RF}(t)$ supplying power $P \propto \int_{-T/2}^{T/2} dt V_{\rm RF}^2(t)$ where $T$ is the oscillation period. Whether the ac power is efficiently transferred to the junction from this ac source ought to depend on $\Xi(\omega)$. We proceed to formally express the explicit form of the effective alternating voltage at the junction. 

This can be done by substituting $\Delta \phi_{\rm x}(t) = \Delta \phi_{\rm x}(t-0^+) = \int_{0^+}^{t} V_{\rm x}(\tau) d\tau$ in eq. (\ref{current_sum_eq}) where $V_{\rm x} = V + V_{\rm RF}(t)$ and $V_{\rm RF}(t)$ is an alternating voltage corresponding to the effect of the RF field given by
\begin{subequations}\label{V_RF_eq}
\begin{align}
V_{\rm RF}(t)=\int_{-\infty}^{+\infty}V_{\rm RF}(\omega)\exp(-i\omega t)d\omega = V_{\rm ac}\cos \Omega t\\
V_{\rm RF}(\omega) =  \frac{V_{\rm ac}}{2}\left \{\delta(\Omega-\omega)+\delta(\Omega+\omega) \right \} 
\end{align}
\end{subequations}
where $V_{\rm RF}(\omega)$, $V_{\rm ac}$ and $\Omega$ is the spectrum, the amplitude and frequency of the RF field respectively and the significance of the $+$ sign is that $0^+$ is experimentally, not computationally equal to $0$; it is the limit $0^+ \equiv t \rightarrow 0$. 

Proceeding, the applied power $P$ of the RF field is the mean-square value of $V_{\rm RF}(t)$ given by, 
\begin{equation}
P \propto \frac{1}{2\pi/\Omega}\int_{-\pi/\Omega}^{\pi/\Omega} [V_{\rm RF}(t)]^2 dt = \frac{V_{\rm ac}^2}{2},  
\label{P_eq}
\end{equation}
where the proportionality factor is the admittance of the junction. 

\subsubsection{Lehmann Weight and Linear Response}

However, we are interested in the power absorbed by the junction $P_{\rm J}$ instead since it modifies the $I-V$ characteristics. In section \ref{Vacuum Excitation}, we argued that this power $P_{\rm J}$ is proportional to $|\Xi(\Omega)|^2$ in the presence of bosonic, anyonic or fermionic excitations.

Within the context of linear response theory\cite{Kubo1957}, this means that the applied voltage $V_{\rm RF}(t)$ acts as an external force, and the effective voltage as a linear
response $V_{\rm RF}'(t)$ of the circuit,
\begin{subequations}
\begin{align}
V_{\rm RF}'(t)  = \int_{-\infty}^{t} \chi(t - s) V_{\rm RF}(s)ds,\\
V'(\omega) = \Xi(\omega)V_{\rm RF}(\omega)\\
\Xi(\omega) = \int_0^{+\infty}\chi(t)\exp (i\omega t) dt,
\end{align}
\end{subequations}
where $\chi(t-s)$ is the response function, making $\Xi(\omega)$ the susceptibility. For the special case discussed in eq. (\ref{eta_eq}), we have $\chi(t) = (1/RC)\exp(-t/RC)$.
Thus, the RF spectrum above gets modified to \begin{subequations}
\label{V_dash_RF_eq}
\begin{align}
V'_{\rm RF}(\omega) =  \frac{V_{\rm ac}}{2}\Xi(\omega)\left \{\delta(\Omega-\omega)+\delta(\Omega+\omega) \right \},
\end{align}
\begin{multline}
V'_{\rm RF}(t)=\int_{-\infty}^{+\infty}V'_{\rm RF}(\omega)\exp(-i\omega t)d\omega,\\ 
= |\Xi(\Omega)|V_{\rm ac}\cos (\Omega t + \eta),
\end{multline}
\end{subequations}
and $P_{\rm J}$ is given by 
\begin{equation}
P_{\rm J} \propto \frac{1}{2\pi/\Omega} \int_{0}^{2\pi/\Omega} [V'_{\rm RF}(t)]^2 dt = \frac{V_{\rm ac}^2}{2}|\Xi(\Omega)|^2 \propto P|\Xi(\Omega)|^2, 
\label{P_eff_eq}
\end{equation}
where we have used eq. (\ref{V_RF_eq}), (\ref{P_eq}) and (\ref{V_dash_RF_eq}), as expected.

\subsubsection{\label{Bloch_section} Unitarity and the topological potential}

\begin{figure}[b]
\begin{center}
\includegraphics[width=\columnwidth]{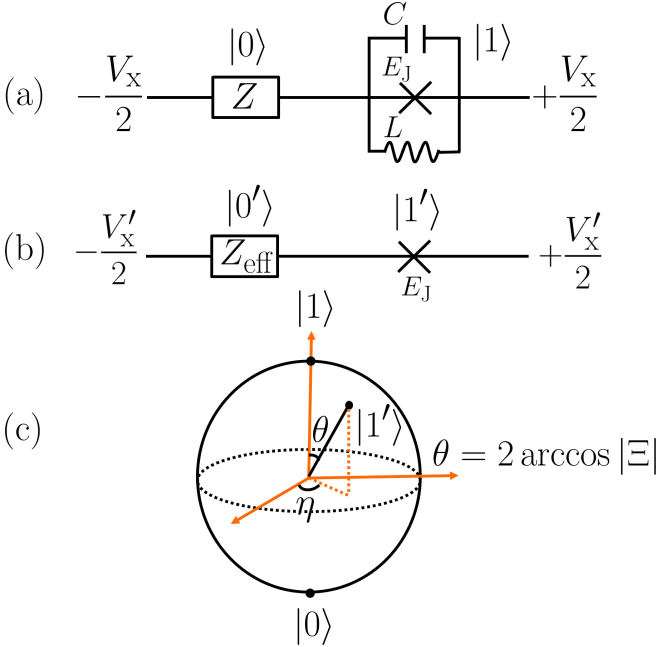}
  \caption{Diagrammatic representation of the unitary transformation implemented by the matrix $\mathcal{U}_{\rm RF}(t = 0)$ in eq. (\ref{unitary_eq}). (a) The equivalent circuit of the Josephson junction labeled by the coupling energy $E_{\rm J}$, the capacitance $C$ and inductance $1/L = \sum_n 1/L_n$ where the admittance $y(\omega) = i\omega C + \sum_n 1/ i\omega L_n$. The junction is coupled serially to the environmental impedance $Z(\omega)$ and symmetrically biased by an external voltage $V_{\rm x}$ where the quantum states of the environment and the junction can be represented by $| 0 \rangle$ and $| 1 \rangle$ respectively; (b) The equivalent circuit of the Josephson junction and its environment. The bias voltage, the quantum states and the environmental impedance are all renormalized by the unitary transformation given by $\mathcal{U}(t)$ in eq. (\ref{unitary_eq}); (c) A Bloch sphere representing the action of the unitary transformation given by $\mathcal{U}(t = 0)$ in eq. (\ref{unitary_eq}), where $\eta$ is the argument of the renormalization factor $\Xi = |\Xi|\exp(i\eta)$ and $\theta = 2\arccos |\Xi|$.}
\label{unitarity_fig}
\end{center}
\end{figure}

This section displays the unitary nature of the renormalization effect. In particular, we show that the renormalization effect can be split into two: The fraction of ac voltage drop at the junction and the fraction of ac drop at the environment. This entails taking the quantum states of the environment and the junction as two orthornormal states $\psi_{\rm J} = |1 \rangle$ and $\psi_{\rm z} = |0 \rangle$ respectively (Fig. \ref{unitarity_fig}) undergoing a time dependent unitary transformation. 

In particular, defining matrices $\mathcal{V}_{\rm x}$ and $\mathcal{U}_{\rm RF}$ and two quantum states $\psi_{\rm J}$ and $\psi_{\rm z}$ of the junction and the environment respectively, \begin{subequations}\label{unitary_eq}
\begin{align}
\mathcal{V}_{\rm x}= 
\frac{1}{2}\left \{ V\sigma_0 + V_{\rm ac}\mathcal{U}_{\rm RF} \right\},\\
\mathcal{U}_{\rm RF}(t)
= \begin{pmatrix}
 \Xi(\Omega)e^{i\Omega t} & -\sqrt{1-|\Xi(\Omega)|^2 }e^{i\Omega t},\\  \sqrt{1-|\Xi(-\Omega)|^2}e^{-i\Omega t} & \Xi(-\Omega)e^{-i\Omega t} \end{pmatrix},
\end{align}
such that,
\begin{align}
\binom{\psi'_{\rm {J}}}{\psi'_{\rm {Z}}}
= \mathcal{U}_{\rm RF} \binom{\psi_{\rm {J}}}{\psi_{\rm {Z}}},\\
V'_{\rm x} = {\rm tr}\left \{ \mathcal{V}_{\rm x} \right \} = V + \int_{-\infty}^{+\infty}V'_{\rm RF}(\omega)\exp(-i\omega t)d\omega,\\
\det\left \{ \mathcal{U}_{\rm RF} \right \} = 1,\\
\mathcal{U}^{\dagger}_{\rm RF}\mathcal{U}_{\rm RF} = \mathcal{U}_{\rm RF}\mathcal{U}^{\dagger}_{\rm RF} = 1,
\end{align}
\end{subequations}
where $\sigma_{0}$ is the $2 \times 2$ identity matrix, we find that these states $\psi_{\rm J} = | 1 \rangle$, $\psi'_{\rm J} = | 1' \rangle$ and $\psi_{\rm Z} = |0\rangle$, $\psi'_{\rm Z} = |0'\rangle$ are normalized $\langle 0 | 0 \rangle = \langle 0' | 0' \rangle = \langle 1 | 1 \rangle = \langle 1' | 1' \rangle = 1$  and orthogonal to each other, $\langle 0 | 1 \rangle = \langle 0' | 1' \rangle = 0$, while orthogonality is preserved under the unitary transformation $\mathcal{U}_{\rm RF}$ leading to a renormalized external voltage $V_{\rm x} = V + V_{\rm ac}\cos(\Omega t) \rightarrow V'_{\rm x} = {\rm tr}\left \{ \mathcal{V}_{\rm x} \right \} = V_{\rm x} + A(t)$. This requires the topological flux $\Delta \Phi(t) \neq 0$ in eq. (\ref{calculation2_eq}) not vanish in the presence of oscillating electromagnetic fields. Solving for the topological potential $A(t)$, we find
\begin{subequations}\label{A_eq}
\begin{align}
A(t) = V'_{\rm x}-V_{\rm x}=V_{\rm ac}\left \{ v(\Omega)\sin\Omega t - u(\Omega)\cos\Omega t\right \}\\
\Gamma(\Omega) = u(\Omega) + iv(\Omega)\\
\Xi(\Omega) = 1-\Gamma(\Omega) = 1 - u(\Omega) - iv(\Omega)
\end{align}
\end{subequations}
with eq. (\ref{A_eq}c) relating the impedance Lehmann weight $\Xi(\Omega)$ to the topological potential amplitude factors $u(\Omega), v(\Omega)$. 

Thus, the purpose of the topological potential is to implement the renormalization scheme above. By eq. (\ref{A_eq}), we find that the topological flux $\Phi(t) = \int_{-\infty}^{t} A(\tau) d\tau$ is ill-defined for $\tau = -\infty$ since $\sin(\omega \infty)$ and $\cos(\omega \infty)$ both oscillate rapidly without converging. Nonetheless, this poses no problem since it is the flux difference $\Delta \Phi(t) = \int_{0}^{t} A(\tau) d\tau$ that appears in the correlation function in eq. (\ref{calculation2_eq}) rendering the $I$--$V$ characteristics in eq. (\ref{IV_eq}) perfectly well-defined. 

Moreover, by eq. (\ref{amplitude_eq}), we find that
\begin{align}
\Xi(\Omega) = \frac{y^{-1}(\Omega)}{y^{-1}(\Omega) + z(\Omega)} = \frac{1}{1 + z(\Omega)y(\Omega)}\\
\Gamma(\Omega) = \frac{z(\Omega)}{y^{-1}(\Omega) + z(\Omega)} = y(\Omega)Z_{\rm eff}(\Omega) 
\end{align}
are ratios of impedances. Thus, in the simple model in eq. (\ref{RCSJ_eq}), power renormalization is negligible ($\Xi(\Omega) \simeq 1$) only for extremely low frequencies satisfying $1/RC \gg \Omega$. However, for samples exhibiting Coulomb blockade that satisfy the Lorentzian-delta function approximation ${\rm Re}\left \{ Z_{\rm eff} \right \} = R/(1 + \Omega^2C^2R^2 ) \sim \pi C^{-1}\delta(\Omega)$, the conductance $R^{-1}$ is extremely small ($1/RC \ll \Omega$) and thus we should expect power renormalization for virtually all applied frequencies.
\\

\subsection{Current--Voltage Characteristics with finite RF Field}

Now that we have the form of the voltage $V'_{\rm x} = V + V_{\rm RF} + A = V + |\Xi|V_{\rm ac}\cos(\Omega t + \eta)$, where $|\Xi|V_{\rm ac} = V_{\rm ac}^{\rm eff}$, it should be substituted into eq. (\ref{IV_eq}) to determine the Cooper-pair and quasi-particle tunneling current in the presence of microwaves. Thus, substituting $V'_{\rm x}$ into eq.  (\ref{IV_eq}) and using the identities, $\sin (x\sin y) = \sum_{n = -\infty}^{\infty} J_{n}(x)\sin ny$ and $\cos (x\sin y) = \sum_{n = -\infty}^{n = \infty} J_{n}(x)\cos ny$ where $J_{n}(x) = (-1)^{-n}J_{-n}(x) = \frac{1}{2\pi}\int ds\exp i(x\sin s - ns)$ is the Bessel function of the first kind, $x,y$ are arbitrary functions and $n \in \mathbb{Z}$ is an integer, the $I$--$V$ characteristics of the irradiated junction can be expressed in terms of the $I$-$V$ characteristics $I_1$ and $I_2$ of the unirradiated junction,
\begin{multline}
I(V) = \sum_{n=-\infty}^{\infty}J_n^2\left(\frac{eV_{\rm ac}^{\rm eff}}{\Omega}\right)\,I_{1}\left(V-\frac{n\Omega}{e}\right)\\
+ \sum_{n=-\infty}^{\infty}J_n^2\left(\frac{2eV_{\rm ac}^{\rm eff}}{\Omega}\right)\,I_{2}\left(V-\frac{n\Omega}{2e}\right).
\label{Tien-Gordon_eq}
\end{multline}
Here, $I_{1,2}$ are the quasi-particle, Cooper pair RF-free $I-V$ characteristics given in eq. (\ref{IV_eq}), $J_n(x)$ are Bessel functions of the first kind where the order $n$ is the number of actual photons absorbed by the junction, $V_{\rm ac}$ is the amplitude of the alternating voltage, $\Omega$ is the energy quantum of individual photons, $V$ and $I$ respectively are the applied dc voltage and tunneling current of the junction. The total current is shifted by the number of photons reflecting energy conservation and is proportional to the square of the Bessel function reflecting the modification of density of states. This equation neglects higher harmonics derived in ref. \citenum{Grabert2015} which were shown to be largely suppressed.\footnote{The amplitude of the higher (current) harmonics has been shown by Grabert in ref. \citenum{Grabert2015}, to correspond to $|\Xi(n\Omega)| = |Z_{\rm eff}(n\Omega)/Z(n\Omega)|$ where $n$ is the $n$-th harmonic (assumed to be positive). Using the impedances $Z(\Omega) =  R$ and $Z_{\rm eff}(\Omega) = 1/(R^{-1} + i\Omega C + 1/i\Omega L)$, the renormalization factor $|\Xi(n\Omega)|$ is a rapidly decreasing function of $n$.} Finally, the minimal conditions for the universality of eq. (\ref{Tien-Gordon_eq}), extending it to a large family of models and beyond the periodic drives, has been systematically given in refs. \citenum{safi2019, safi2014, safi2010}. 

\section{\label{Sec: Arrays} Renormalization of Applied Oscillating Voltages in Linear Arrays of Josephson Junctions}

\subsection{Renormalization factor and soliton length}

We consider the action for $N_{0}$ number of Josephson junction elements where the action resembles that for a single junction given in eq. (\ref{S_eff_CL_eq}) where all the admittance elements in the expression are replaced by $(N_{0}-1)\times(N_{0}-1)$ matrices,
\begin{subequations}
\begin{widetext}
\begin{align}
S^{\rm A}_{\rm z} = \sum_{j,k = 1}^{N_{0} -1}\int dt \left \{ \frac{1}{2e^2}C_{jk}\frac{\partial\phi'_{j}}{\partial t}\frac{\partial\phi'_{k}}{\partial t} - \frac{1}{4\pi e^2}\int ds\,\phi_{j}'(s)\left [ \frac{\partial (Z^{-1})_{jk}(t-s)}{\partial t} \right ]\phi_{k}'(t)\right \} - E_{\rm J}\sum_{i = 1}^{N_{0} - 1}\int dt\,\cos(\phi'_{i}(t) - \phi'_{i + 1}(t)), 
\label{Action_Array_eq}
\end{align}
\end{widetext}
where $C_{jk} = (C_{0} + 2C)\delta_{j,k} - C\delta_{j + 1, k} - C\delta_{j - 1, k}$ is the capacitance matrix of the array with Josephson junctions of equal capacitance $C$, $N_{0} - 1$ is the number of islands, $C_{0}$ is the stray capacitance of each island, $\phi'_{j}$ and $Q'_{j} = e^{-1}\sum_{k}C_{jk}\partial\phi'_{k}/\partial t$ the phase and charge of each island respectively, $E_{\rm J} \ll e^2/2C$ the Josephson coupling energy of each island and $(Z^{-1})_{jk}$ is the unspecified environment admittance matrix of the array.\footnote{This action has been considered before within the context of one dimensional XY model of topological phase transitions, e.g. in ref. \citenum{JJ_Array_XY} with $(Z^{-1})_{ij} = 0$. This reference can be consulted for introduction on how to approach dynamics and phase transitions in such a linear array.} 
\begin{figure}[!t]
\begin{center}
\includegraphics[scale = 1]
{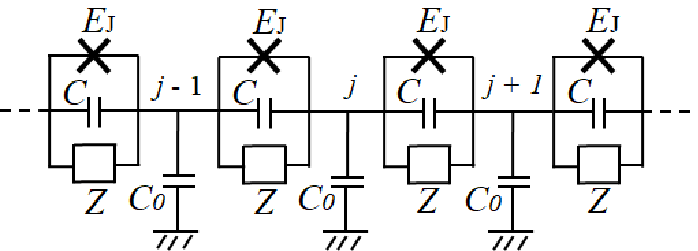}
  \caption{Chain of Josephson junction arrays representing the locations of each element in the array. Here, $C$ is the junction capacitance, $Z$ is the junction environmental impedance and $E_{\rm J}$ is the Josephson coupling energy. Each island is labeled by an index $j$ and is characterized by a self-capacitance $C_{0}$. The action for this circuit is given by eq. (\ref{Action_Array_eq}) and eq. (\ref{Action_Array_Fourier_eq})}
\label{JJ_Array_Chain}
\end{center}
\end{figure}
The Fourier transform of this action with $E_{\rm J} \rightarrow 0$ as a perturbation parameter takes the general form, 
\begin{align}
S_{\rm z}^{\rm A} = \frac{2\pi}{2e^2}\sum_{j,k = 1}^{N_{0} - 1}\int d\omega\phi'_{j}(\omega)i\omega(Z^{\,-1}_{\rm eff})_{jk}(\omega)\phi'_{k}(-\omega),
\label{Action_Array_Fourier_eq}
\end{align}
\end{subequations}
where $(Z^{-1}_{\rm eff})_{jk} = i\omega C_{jk} + (Z^{-1})_{jk}(\omega)$ is the effective admittance matrix. It is now straightforward to determine the phase-phase correlation function $\langle \phi'_{j}(t)\phi'_{k}(t) \rangle$ by recalling that its Fourier transform is equal to the imaginary part of the Green's function of the action read-off from eq. (\ref{Action_Array_Fourier_eq}) to yield $G_{jk}^{\rm A}(\omega) = 2e^2i\omega^{-1}(Z_{\rm eff})_{jk}(\omega)$ where $(Z_{\rm eff})_{jk}(\omega)$ is the inverse matrix of $(Z^{-1}_{\rm eff})_{jk}$. This procedure yields, $\int dt\,\langle \phi'_{j}(t)\phi'_{k}(0) \rangle \exp(i\omega t) = G_{jk}^{\rm A}(\omega) - G_{jk}^{\rm A}(-\omega) = 2e^2\omega^{-1}\left \{(Z_{\rm eff})_{jk}(\omega) + n.f.\right \}$. This suggests that the Lehmann/wavefuntion renormalization weight is a matrix of the form, $\sum_{l}(Z_{\rm eff})_{jl}(Z^{-1})_{lk} = \Xi_{jk}(\omega)$. The amplitude of the applied ac voltage will be modified by its determinant $\Xi_{\rm A} = \det(\Xi_{ij}(\omega))$.  

\begin{figure}[!t]
\begin{center}
\includegraphics[scale = 1]
{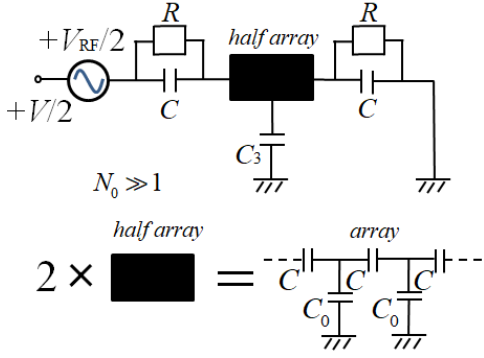}
  \caption{Simplified circuit of a symmetrically biased infinite array with $N_{0} \gg 1$ showing only half the array as a black box where the effective capacitance of the whole array is given by $2C_{3} = \frac{1}{2}\left ( C_{0} + \sqrt{C^2 + 4CC_{0}}\right )$.}
\label{2JJ_Array}
\end{center}
\end{figure}

For Gaussian correlated phases, $\langle \phi'_{j}(t)\phi_{k}'(s) \rangle = 0$ with $k \neq j$, the impedance matrix $(Z_{\rm eff})_{jk}(\omega)$ has to be diagonalized, with $(Z_{\rm eff})_{jk}(\omega) = 0$ for $j \neq k$. This is akin to setting all the phase-phase interaction terms to zero. However, this is not the case since the islands will effectively interact when a charge soliton propagates along the array constituting a current. The injection of a soliton/anti-soliton pair into the array depends on the electrostatic potential at the junction at the edge and the one at the center of the array labeled $1$ and $2$. We shall approximate the array as infinite with $N_{0} \gg 1$ junctions, where each junction has a capacitance $C$ and environmental impedance $Z(\omega) = R$, leading to the effective circuit depicted in Fig. \ref{2JJ_Array}. Thus, the capacitance of the rest of the array is computed by recognizing that for an infinite array, neglecting the capacitance of the first junction $C$ and the self-capacitance of the first island $C_{0}$ does not alter the capacitance $C_{\rm r}$ of the rest of the array,
$C_{\rm r}^{-1} = C^{-1} + (C_{0} + C_{\rm r})^{-1}$. Solving for $C_{\rm r}$, we find
\begin{multline*}
\frac{1}{(C_{0} + C_{\rm r} + C)C_{\rm r}} = \frac{1}{(C_{0} + C_{\rm r})C}\\
\rightarrow C_{\rm r}^{2} + C_{0}C_{\rm r} - C_{0}C = 0 \rightarrow C_{\rm r} = \frac{1}{2}\left (-C_{0} + \sqrt{C^2_{0} + 4CC_{0}} \right ).
\end{multline*}
The total capacitance of the infinite array $C_{\rm A}$ (excluding the first junction) is given by,
\begin{align*}
C_{\rm A} = C_{0} + C_{\rm r} = \frac{1}{2}\left ( C_{0} + \sqrt{C^2_{0} + 4CC_{0}} \right ).  
\end{align*}
We thus set the capacitance of half the array as $C_3 = C_{\rm A}/2$.

Taking the junction phases as $\phi_{1}$ and $\phi_{2}$, we can write the exact action of the effective circuit as,
\begin{multline}\label{Action_Array_eff_eq}
    S_{\rm eff}^{\rm A} = \frac{C}{2e^2}\int dt\left \{\left ( \frac{\partial \phi_1}{\partial t} \right )^2 + \left ( \frac{\partial \phi_2}{\partial t} \right )^2\right \}\\
    + \int dt \frac{C_{3}}{2e^2}\left ( \frac{\partial \phi_1}{\partial t} - \frac{\partial \phi_2}{\partial t} \right )^2\\
    -\frac{1}{4\pi e^2}\int dtds\,\phi_{1}(t)\frac{\partial Z^{-1}(t - s)}{\partial t}\phi_{1}(s) \\
    - \frac{1}{4\pi e^2}\int dtds\,\phi_{2}(t)\frac{\partial Z^{-1}(t - s)}{\partial t}\phi_{2}(s).  
\end{multline}
Note that, when Fig. (\ref{2JJ_Array}) instead of Fig. (\ref{JJ_Array_Chain}) is used to evaluate the effective impedance of the array, any possible corrections to eq. (\ref{Action_Array_eff_eq}) ought to be negligible. Such corrective terms in eq. (\ref{Action_Array_eff_eq}) ought to be small but finite if and only if one perturbatively moves from the $N_0 \gg 1$ regime of the infinite array towards the $N_0 > 1$ regime corresponding to eq. (\ref{Action_Array_eq}). 
Thus, eq. (\ref{Action_Array_eff_eq}) leads to the following $2 \times 2$ capacitance and admittance matrices respectively, 
\begin{align*}
i\omega C_{jk} = i\omega \begin{pmatrix}
C + C_3 & -C_3\\ 
-C_3 & C + C_3
\end{pmatrix},\,\,\,
(Z^{-1})_{jk} = \frac{1}{R} 
\begin{pmatrix}
1 & 0\\ 
0 & 1
\end{pmatrix},
\end{align*}
where $j,k = 1,2$. The effective admittance matrix is thus given by their sum, $(Z^{-1}_{\rm eff})_{jk} = (Z^{-1})_{jk} + i\omega C_{jk}$. 

Thus, the array via the interaction term with $C_{3} = C_{\rm A}/2$ modifies the phase-phase correlations by destroying their Gaussian nature as the impedance matrix is shifted as $(i\omega C + 1/R)\delta_{ij} \rightarrow (Z^{-1}_{\rm eff})_{jk}$. In turn, this corresponds to a Lehmann weight,
$\Xi_{jk} = \sum_{l = 1}^{2}(i\omega C + 1/R)\delta_{jl}(Z_{\rm eff})_{lk} = (i\omega C + 1/R)(Z_{\rm eff})_{jk}$. This will affect the amplitude of the applied oscillating voltage by a Lehmann weight $\det(\Xi_{jk})$. When the angular frequency $\omega$ is much larger than the inverse of the time constant $RC$, we find,
\begin{align}
    \Xi_{\rm A} = \lim_{\omega RC \rightarrow \infty} \det(\Xi_{jk}) = \frac{C}{C + C_{\rm A}} = \exp(-\Lambda^{-1}),
\end{align}
where $\Lambda$ is the soliton length of the array. When an alternating voltage is applied across the array, the amplitude of the oscillating voltage will be renormalized by $\Xi_{\rm A} \sim \exp(-\Lambda^{-1})$. This represents a damping of the applied power of applied oscillating voltage. 

\subsection{\label{Effective field theory}  Soliton Field Theory Origin of the Lehmann Weight in an Infinite Array}

Consider the charge soliton lagrangian of the array,
\begin{subequations}
\begin{equation}
\mathcal{L}_{\rm sol} = \frac{1}{2\pi}\int dx\left \{ \frac{1}{2}\left ( \frac{\partial \chi}{\partial t}  \right )^2 - \frac{1}{2}\left ( \frac{\partial \chi}{\partial x}  \right )^2 - \frac{2}{\Lambda^2}\sin^2(\chi/2) \right \},
\label{soliton_eq}
\end{equation}
where the co-ordinates $x \equiv x/a$, $t \equiv v_{0}t/a$ are dimensionless, and $a = 1$ is the length of the islands (lattice constant), $v_{0} = 1$ is the velocity of electromagnetic radiation along $x$. The Euler-Lagrange equations of eq.  (\ref{soliton_eq}) yield the solution, 
\begin{align}
\chi = 4\arctan\exp\left \{ \Lambda^{-1}\gamma(x \pm vt)\right \} + 2\pi n,
\label{sol_soliton_eq}
\end{align}
\end{subequations}
with $\gamma = 1/\sqrt{1-v^2}$ the Lorentz factor. 
Plugging in eq.  (\ref{sol_soliton_eq}) into eq.  (\ref{soliton_eq}), we can eliminate $\partial/\partial t$ in favor of $\partial/\partial x$,
\begin{subequations}
\begin{align}
 \left ( \frac{\partial \chi}{\partial t}  \right )^2 = v^2\left ( \frac{\partial \chi}{\partial x}  \right )^2\\
 \mathcal{L}_{\rm sol} = \frac{-1}{\pi}\int dx\left \{\frac{1}{4\gamma^2}\left ( \frac{\partial \chi}{\partial x}  \right )^2 + \Lambda^{-2}\sin^2(\chi/2) \right \}.
\end{align}
Observing that since the integrand is quadratic, we can apply the Bogomol'nyi inequality\cite{Zee2010} ($A^2 + B^2 \geq 2|AB|$) to evaluate the mass, $M$ given by,
\begin{multline}
\mathcal{L}_{\rm sol} \geq M = \frac{1}{\pi}\int dx\left | \frac{1}{2\gamma \Lambda}\left ( \frac{\partial \chi}{\partial x}  \right )\sin(\chi/2) \right |\\
= \frac{1}{\gamma \Lambda \pi}\left |\int dx\frac{\partial \cos(\chi/2)}{\partial x} \right |
= \frac{1}{\gamma \Lambda \pi}\left | \left [ \cos(\chi/2) \right ]_{-\infty}^{+\infty} \right |\\
= \frac{1}{\gamma \Lambda \pi}\left | \cos(\pi) - \cos(0) \right | = \frac{2}{\gamma \Lambda \pi} \equiv \frac{E_{0}}{\gamma \Lambda}
\label{Bogomol'nyi_eq}
\end{multline}
\end{subequations}
with $E_{0} = 1/a\pi$. Treating the solitons as charged dust of mass density, $M/\mathcal{V}$ where $u^{\mu} = dx^{\mu}/d\tau = (\gamma, \pm v\gamma, 0, 0)$ is the four-velocity, $\mathcal{V} = \mathcal{A}l$ is the volume and $l$ the length of the array, we can introduce the energy-momentum tensor,
\begin{align}
T^{\mu\nu} = \frac{2M}{\mathcal{V}}u^{\mu}u{^\nu} + \varepsilon_{0}\varepsilon_{\rm r}\left \{F^{\mu\sigma}F^{\nu}_{\,\,\sigma} - \frac{1}{4}F^{\sigma\rho}F_{\sigma\rho}\eta^{\mu\nu}\right \}.
\end{align}
The total energy is given by,
\begin{subequations}
\begin{multline}
E =  \int_{\mathcal{V}} d\,^3x \langle T^{00} \rangle = \frac{1}{\mathcal{V}} \int_{\mathcal{V}} d\,^3x \langle  2M u^0u^0 \rangle\\
+ \frac{1}{\mathcal{V}} \sum_{s = \pm 1}\int_{\mathcal{V}} d\,^3x\, \omega \left ( \langle a_{s}(\omega)a_{s}(-\omega) \rangle_{\omega} + \frac{{\rm sgn}(\omega)}{2} \right ),
\end{multline}
with $s = \pm 1$ the photon polarization states. Note that,
\begin{align}
\langle M u^0u^0 \rangle = \frac{E_0}{\Lambda}\langle \gamma \rangle  \simeq \frac{E_0}{\Lambda} + \frac{E_0}{\Lambda} \frac{\langle v^2\rangle }{2} =  \frac{1}{\Lambda}\left ( E_0 + \frac{1}{2\beta} \right ),
\end{align}
\label{free_energy_sol_eq}  
\end{subequations}
where,
\begin{align*}
    \langle \cdots \rangle \equiv \left ( \int_{-\infty}^{+\infty} dv\, \exp -\beta E_0v^2/2 \right )^{-1}\int_{-\infty}^{+\infty} dv\, (\cdots \exp -\beta E_0v^2/2),
\end{align*}
is the Boltzmann average for a gas of (anti-)solitons in 1 + 1 dimensions. Comparing eq.  (\ref{free_energy_sol_eq}) to eq.  (\ref{free_energy_eq}) (neglecting the vacuum evergy $E_0/\Lambda$ and considering only one photon polarization mode), we conclude that the array is an effective single junction with $\Xi_{\rm A} = \exp(-\Lambda^{-1})\exp(2\pi m i)$. 

\section{Application: Optimization of linear arrays for classical RF field power detection}

We treat the array of Josephson junctions as an effective single junction based on the arguments presented in Appendix \ref{PP_Approach}. This entails using the $I$-$V$ characteristics given in eq. (\ref{I_V_array_eq}) under irradiation of the RF field, together with the renormalized external voltage $V'_{\rm x}(t) = V + V_{\rm ac}^{\rm eff}\cos(\Omega t + \eta)$ in eq. (\ref{I_V_array_eq}), which simply results in eq. (\ref{Tien-Gordon_eq}), 
\begin{align}
   I_{\rm A}(V) = \sum_{\kappa = 1}^{2}\sum_{n=-\infty}^{\infty}J_n^2\left(\frac{\kappa eV_{\rm ac}^{\rm eff}}{\Omega}\right)\,I_{\kappa}\left(V-\frac{n\Omega}{\kappa e}\right),
\label{Tien-Gordon_eq2}
\end{align}
where $V_{\rm ac}^{\rm eff} = |\Xi(\Omega)|V_{\rm ac}$, $I_{\kappa}(V)$ are the $I$-$V$ characteristics of the array with $V_{\rm ac} = 0$ and the $P_{\kappa}(E) = \frac{1}{2\pi}\int dt P_{\kappa}(t) \exp(iEt)$ functions appearing in $I_{1}, I_{2}$ in eq. (\ref{IV_eq}) are rescaled as $P_{\kappa}(t) \rightarrow [P_{\kappa}(t)]^{\langle N_{0} \rangle}$. This is the celebrated Tien-Gordon equation\cite{Tien-Gordon1963} describing photon-assisted tunneling of Cooper-pairs or quasi-particles of charge $\kappa e$ across a barrier. However, in the case of the array, current is predominantly generated by charge solitons. Thus, eq. (\ref{Tien-Gordon_eq2}) ought to correspond to charge solitons/anti-solitons injected into the array by the influence of the RF field. 

The universality of eq. (\ref{Tien-Gordon_eq}) and hence eq. (\ref{Tien-Gordon_eq2}) is apparent when we observe that the necessary and sufficient conditions for reproducing it are: 
\begin{enumerate}
    \item The unirradiated characteristics of the sample take the form $I_{\kappa}(V) = ie\int_{-\infty}^{+\infty} dt\, \alpha_{\kappa}(t)P_{\kappa}(t)\sin\left [\kappa e\int_{0^+}^{t} dt'V \right ]$, where the structure coefficients $\alpha_{\kappa}(t)$ depend on the density of states of the tunneling particles;
    \item The effect of the RF field be to shift $V$ to $V'_{\rm x}(t) = V + V_{\rm ac}^{\rm eff}\cos(\Omega t + \eta)$.
\end{enumerate}
Thus, the aforementioned rescaling of $P_{\kappa}(t)$ in the case of the array merely corresponds to a modification of the structure coefficients $\alpha_{\kappa}(t)$ in condition 1 by a multiplicative factor $[P_{\kappa}(t)]^{\langle N_{0} \rangle - 1}$. Moreover, the renormalization effect highlighted in this paper concerns only condition 2. As a consequence, it is sufficient to work in the classical limit $\Omega \ll \kappa eV_{\rm ac}$ of the RF field, corresponding to multi-photon absorption by the sample (single junction or array). Setting $\kappa eV_{\rm ac}\,{\rm sin}\,\theta = n\Omega$, the sum over photon number $n$ can be approximated by an integral formula corresponding to the classical detection of radiation,\cite{Hamilton1970}
\begin{equation}
I_{\rm A}(V)=\frac{1}{\pi}\int_{-\pi/2}^{\pi/2}\,I_{0}\left(V-|\Xi(\Omega)|V_{\rm ac}\,{\rm sin}\,\theta\,\right)\,{\rm d}\theta,
\label{Hamilton_eq}
\end{equation}
where $I_0(V) = \sum_{\kappa = 1}^{2} I_{\kappa}(V)$ is the sum of the quasi-particle and Cooper-pair currents. For the simulation, we used the measured characteristics, $I_{0}(V)$ of a linear array of 10 small Josephson junctions exhibiting distinct Coulomb blockade characteristics, thus by-passing simulating condition 1 which merely corresponds to standard $P(E)$ theory.\cite{Ingold_Nazarov1992, ingold1994_PE} 
\begin{table}[!t]
\caption{Average parameters per junction the array of 10 Aluminium ($\rm Al$)/Aluminium Oxide ($\rm Al_xO_y$)/Aluminium ($\rm Al$) Josephson junctions whose measured characteristics, $I_{0}(V)$ have been used in the simulation with eq. (\ref{Hamilton_eq}). The parameters consecutively are, the capacitance $C$, tunnel resistance $R_{\rm T}$, Josephson coupling energy $E_{\rm J}$, charging energy $E_\mathrm{c}$ and $E_{\rm J}/E_{\rm c}$ ratio.}\label{table:parameter}
    \begin{center}
    \scalebox{1.2}{
        \begin{tabular}{ccccc}\hline
            $C$ [fF] & $R_{\rm T}$ [k${\rm \Omega}$] & $E_{\rm J}$ [$\mu$eV] & $E_{\rm c}$ [$\mu$eV] & $E_\mathrm{J}/E_\mathrm{c}$\\
            \hline
            0.8 & 10 & 62.5 & 96 & 0.65\\
            \hline
        \end{tabular}}
     \end{center}
\end{table}

\begin{figure}
\begin{center}
\includegraphics[width=\columnwidth]{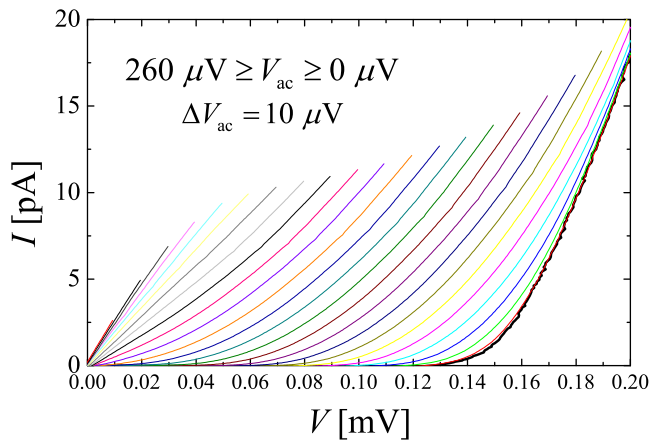}
  \caption{Simulated $I$--$V$ curves using eq. (\ref{Hamilton_eq}) and the measured characteristics $I_{0}(V)$ (black bold curve) of a linear array of 10 Josephson junctions with average parameters per junction given in Table \ref{table:parameter}, for amplitude range $0$ $\mu$V $\leq V_{\rm ac} \leq 260$ $\mu$V, where the interval between adjacent curves is $\Delta V_{\rm ac} = 10$ $\mu$V and the renormalization factor is set to unity, $|\Xi| = 1$. These characteristics simulate the lifting of Coulomb blockade by RF power.}
\label{fig_sim}
\end{center}
\end{figure}
The target parameters of the linear array per junction displayed in Table \ref{table:parameter} were determined by design during electron-beam lithography and oxidation during shadow evaporation. The tunnel resistance $R_{\rm T}$ and $E_{\rm c}$ are calculated respectively from the offset voltage and the differential conductance ${\rm d}I_{0}(V)/{\rm d}V$.\cite{Averin1991, Delsing1992} The differential conductance, alongside the measured $\Delta$--$H$ dependence determine the superconducting gap $\Delta$. The Josephson coupling energy, $E_{\rm J}$ is then determined by the Ambegaokar-Baratoff relation\cite{A-B} using the measured superconducting gap. Note that $E_{\rm J} < E_{\rm c}$ and $R_{\rm T} > R_{\rm Q}/4 \simeq 6.45$ k$\rm \Omega$ satisfy the small junction and Coulomb blockade conditions respectively.

The $I_0(V)$ characteristics of the linear array (given by the black bold curve in Fig. \ref{fig_sim}) was measured via the well-known $r$-bias method.\cite{Delsing_thesis} It entails incorporating a fixed resistance given by $r =$ 1 M${\rm \Omega}$ or 5 M${\rm \Omega}$ serially connected to the array and biasing both (the resistor and the array) with a voltage, and measuring the current and voltage values employing differential amplifiers with high input impedance. Noise reduction was achieved by applying half the dc voltage in each terminal with opposite polarity ($-V/2$ and $V/2$) relative to the ground. 

The simulated curves given in Fig. \ref{fig_sim} were numerically produced using eq. (\ref{Hamilton_eq}), by applying Simpson's rule to approximate the integral,
\begin{align*}
I_{\rm A} \simeq \frac{\Delta \theta}{3\pi}\left [ I_{0}(\theta_0) + I_{0}(\theta_{2l})
+ 2\sum_{m = 1}^{l}I_{0}(\theta_{2m}) 
+ 4\sum_{m = 1}^{l}I_{0}(\theta_{2m+1})\right ]
\end{align*}
where $\theta_m = -\pi/2 + m\pi/2l$ defined between the entire integration interval $\pi = 2l\Delta\theta$ for $2l + 1$ values bounding $2l$ equally spaced intervals of width $\Delta\theta$. The sum was then carried out using spline interpolation.\cite{Spline} The results of the simulation have been plotted in Fig. \ref{fig_sim} for the amplitude range $0$ $\mu$V $\leq V_{\rm ac} \leq 260$ $\mu$V, where the interval between adjacent curves is $\Delta V_{\rm ac} = 10$ $\mu$V and $|\Xi| = 1$. 

Figure \ref{fig_sim} displays the lifting of Coulomb blockade characteristics by the power of the RF field. Such characteristics have been experimentally observed \cite{Delsing1992, Liou2014, Kanyolo2020JLPT} for both quasi-particle and Cooper-pair tunneling. This phenomenon is dual to microwave-enhanced phase diffusion.\cite{Koval2010, Liou2008, Liou2014} The simulated results for $|\Xi| < 1$ exhibit exactly the same Coulomb blockade lifting behaviour. 

\begin{figure}[t!]
\begin{center}
\includegraphics[width=\columnwidth]{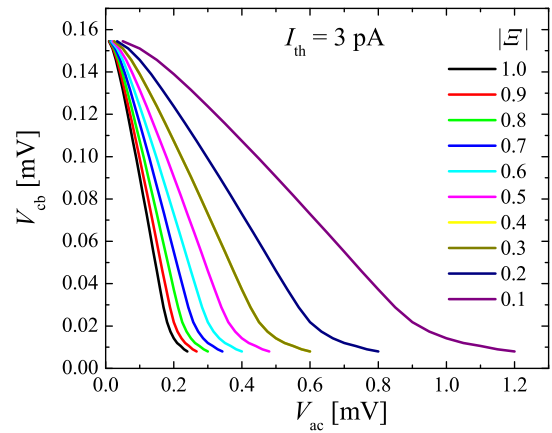}
  \caption{Simulated response of the Coulomb blockade threshold voltage $V_{\rm cb}$ at a threshold current $I_{\rm th} = 3$ pA to RF power $V_{\rm ac}$ for selected values of the renormalization factor in the range $0.1 \leq |\Xi| \leq 1$. These characteristics can be exploited to design an RF power detector with a sensitivity $|{\rm d}V_{\rm cb}/{\rm d}V_{\rm ac}|$ proportional to $|\Xi|$.\cite{Kanyolo2020JLPT}}
\label{Vac_Vcb}
\end{center}
\end{figure}

The lifting of Coulomb blockade characteristics with applied RF power can be exploited to design a microwave power detector by defining the Coulomb blockade threshold voltage $V_{\rm cb}$ at a given threshold current $I_{\rm th}$ and tracking its value in the presence of RF power. To illustrate this, we have plotted the $V_{\rm cb}$--$V_{\rm ac}$ dependence at $I_{\rm th} = 3$ pA in Fig. \ref{Vac_Vcb} for simulated curves using eq. (\ref{Hamilton_eq}) with selected values of the renormalization factor, $|\Xi| \leq 1$. The sensitivity of such as detector is given by $|{\rm d}V_{\rm cb}/{\rm d}V_{\rm ac}| \propto |\Xi|$. However, observe that detection range for $|\Xi| = 1$ given by $0$ mV $\leq V_{\rm ac} \leq 0.26$ mV is significantly smaller than that for $|\Xi| < 1$. For instance, the viable microwave power detection range for $|\Xi| = 0.1$ is $0$ mV $\leq V_{\rm ac} \leq 1.2$ mV. This implies that the optimum microwave detector, depending on use, should lie between $0.1 < |\Xi| < 1$. Since the renormalization factor for a linear array is predicted to be $|\Xi| \sim \exp(-\Lambda^{-1})$, this corresponds to an optimization condition for the soliton length of a suitable linear array for microwave detection. Such a linear array with soliton length $\Lambda \simeq 9$, corresponding to $|\Xi| \sim \exp(-\Lambda^{-1}) \simeq 0.89$, has been experimentally shown to have a sensitivity slightly greater than $10^6$ V/W.\cite{Kanyolo2020JLPT}  

\begin{table*}[!t]
\caption{Electromagnetic quantities and their Rescaled Expressions.}\label{table: renorm}
\begin{center}
\begin{tabular}{ccc} 
\hline
Quantity & Expression & Rescaled Expression\\ 
\hline\hline
Environmental impedance (Single Junction) & $Z(\omega)$ & $Z_{\rm eff}(\omega) =  \Xi(\omega)Z(\omega)$\\ 
Lehmann weight (Single Junction) & $1$ & $\Xi(\omega) = [1 + i\omega CZ(\omega)]^{-1}$\\ 
Microwave amplitude (single junction) & $V_{\rm ac}$ & $V^{\rm eff}_{\rm ac} = |\Xi(\Omega)|V_{\rm ac}$\\ 
Environmental Impedance (long array: Fig. \ref{2JJ_Array}) & $(i\omega C + 1/R)\delta_{jk}$ & $(Z^{-1}_{\rm eff})_{jk} + i\omega C_{jk}$\\ 
Lehmann Weight (long array) & $\det\left [\delta_{jk}\right ] = 1$ & $\Xi_{\rm A} = \det\left [ i\omega C + 1/R)(Z_{\rm eff})_{jk}\right ] \sim \exp(-\Lambda^{-1})$\\ 
Microwave amplitude (long array) & $V_{\rm ac}$ & $V_{\rm ac}^{\rm eff} \sim \exp(-\Lambda^{-1})V_{\rm ac}$\\ 
\hline
\end{tabular}
\end{center}
\end{table*}

\section{Discussion}

\paragraph{Implication}

In the case of the single Josephson junction, the renormalization of the amplitude of applied oscillating electromagnetic fields is implemented by linear response. This entails the excitation of `particles' appearing as a mass gap $M - i\varepsilon_{m} = -\beta^{-1}\ln \Xi(\omega)$ in the thermal radiation spectrum, where $\varepsilon_{m}$ is the Matsubara frequency\cite{Matsubara1955}, $\Xi(\omega) = [1 +  y(\omega)Z(\omega)]^{-1}$ is the linear response function, $Z(\omega)$ is the environmental impedance of the junction and $y(\omega) \simeq i\omega C$ is the impedance of the junction. Likewise, when an infinitely long array\cite{Bakhvalov1989, Delsing1992} is modeled as half the infinite array interacting with two junctions, one at one edge of the the array and the other at the center, as illustrated in Fig. \ref{2JJ_Array}, we find an additional Lehmann weight $\Xi_{\rm A} = \exp(-\Lambda^{-1})$. This requires that applied oscillating electric fields are damped by the same factor over a finite range of electric fields along the array. This is dual to the Meissner effect where the Cooper-pair order parameter leads to a finite range of the magnetic field.

A Josephson junction circuit that exhibits a large Coulomb blockade voltage is ideal for the observation of the renormalization effect. In particular, for the single junction, power renormalization is negligible ($\Xi(\Omega) \simeq 1$) only for extremely low microwave frequencies satisfying $1/RC \gg \Omega$. However, for samples exhibiting Coulomb blockade that also satisfy the Lorentzian-delta function approximation ${\rm Re}\left \{ Z_{\rm eff} \right \} = R/(1 + \Omega^2C^2R^2 ) \sim \pi C^{-1}\delta(\Omega)$, the conductance $1/R$ is extremely small ($1/RC \ll \Omega$) and thus we should expect power renormalization
for virtually all applied frequencies. In the case of long arrays ($N_{0} \gg \Lambda$) with $\Xi \simeq 1$, RF amplitude renormalization should be readily observed due to the additional factor $|\Xi(\Omega)| \sim \exp(-\Lambda^{-1})$. \cite{Kanyolo2020JLPT} 
Finally, a list of the electromagnetic quantities and their rescaled formulae is displayed in TABLE \ref{table: renorm}.

\paragraph{Backaction Considerations}
The form of the admittance $y(\omega)$ given in eq. (\ref{impedance_eq}) neglects the back-action of the Josephson junctions on the environment (with the bath and the junction becoming entangled) which has been reported to dramatically change the predictions of the $P(E)$ theory.\cite{Joyez2013, Back_React2019, E_Jren2019} This back-action manifests through the non-linear inductive response of the junction where the Josephson coupling energy is renormalized, $E_{\rm J}^{\rm eff} = E_{\rm J}\langle \cos(\Delta 2\phi_{\rm J}) \rangle = E_{\rm J}\exp(-\langle \Delta 2\phi_{\rm J}^2 \rangle)$. In our work, we have made an implicit assumption that, whenever the Josephson coupling energy $E_{\rm J}$ is considerably small compared to all relevant energy scales such as the charging energy $E_{\rm c}$, and the current-voltage characteristics of the junction do not exhibit a superconducting branch, this back-reaction can be taken to be small. Nonetheless, considering this back-reaction in our theoretical framework especially in the case of a single Josephson junction is certainly warranted since the back-reaction has been suggested to dramatically alter the superconductor-insulator transition conditions for the Josephson junction.\cite{Back_React2019}

Within our path integral approach, considering such effects entails performing the path integral for the rescaled $P(E)$ function given by $P^{\rm eff}_{\kappa = 2}(E) = \int dt\, P^{\rm eff}_{\kappa = 2}(t)\exp(-iEt)$ where, 
\begin{multline*}
P^{\rm eff}_{\kappa = 2}(t) = \int D\phi_{\rm J} \cos 2\Delta\phi_{\rm J}(t)\exp \left (iS_{\rm CL} \right )\\
\times \exp \left (- i\int dt\,E_{\rm J}\cos 2\phi_{\rm J}(t)\right ),
\end{multline*}
which is challenging to carry out successfully to all orders of perturbation. Typically, the exponent is linearized as $\cos(2\phi_{\rm J}) \simeq 1 - (2\phi_{\rm J})^2/2 + (2\phi_{\rm J})^4/4 + \cdots$ which becomes the $\phi^{4}$ theory.\cite{Peskin_Schroeder1995} In turn, at order $\phi^2$, the renormalized $E_{\rm J}$ is expected to enter the usual Caldeira-Leggett expression as an inductance, $L_{\rm J}^{\rm eff} = 1/(2e)^{2}E_{\rm J}^{\rm eff}$, $P_{\kappa = 2}(E) = \int dt P_{\kappa = 2}(t)\exp(-iEt)$ where, $P_{\kappa = 2}(t) = \int D\phi \cos\Delta\phi(t)\exp \left (iS_{\rm CL}^{\rm eff}\right )$, yields a $P_{\kappa = 2}(E)$ function with a linear inductive part in the impedance as given by $y(\omega)$ in eq. (\ref{impedance_eq}). Further discussion on the experimental and theoretical results of $E_{\rm J}-$based renormalization, the back-action and other possible effects in Shapiro step-based experiments when $E_{\rm J}/E_{\rm c} > 1$ is beyond the scope of this work.

\paragraph{Summary}

We have employed path integral formalism to derive the Cooper-pair current and the BCS quasi-particle current in small Josephson junctions and introduced a model which transforms the infinitely long array\cite{Bakhvalov1989} into an effective circuit with a rescaled environmental impedance, $Z_{\rm eff}^{\rm A} \sim \exp (-\Lambda^{-1})\Xi(\omega)Z(\omega)$, where $Z(\omega)$ is the environmental impedance as seen by a single junction in the array. As is the case for the single junction, we expect that $\Xi(\omega) = \exp [-\beta M(\omega)-\Lambda^{-1}]\exp i\varepsilon_{m}$ also acts as a linear response function for oscillating electromagnetic fields, and can be interpreted as the probability amplitude of exciting a `particle' of mass $M + \beta^{-1}\Lambda^{-1}$ from the junction ground state by the electromagnetic field,\cite{Kanyolo_Berry} with the quantum statistics of this `particle' determined by the complex phase $\varepsilon_{m}$ identified as the Matsubara frequency.\cite{Matsubara1955} In the case of the infinite array, this `particle' corresponds to a bosonic charge soliton injected into the array.\footnote{Appendix \ref{Effective field theory}} This analysis does not take into account random offset charges which are known to act as static or dynamical background charges in the islands of the array, resulting in shifting of the threshold voltage $V_{\rm th}$ and noise generation affecting the soliton flow along the array.\cite{cole2014, walker2015}

\paragraph{Application}
Since the quasi-particle current naturally reduces to the normal current and the supercurrent vanishes when the superconducting gap goes to $\Delta = 0$, the final expression of the tunnel current eq. (\ref{Tien-Gordon_eq}) is essentially the time averaged current result previously proposed in ref. \citenum{Grabert2015}. In the classical limit when the RF frequency $\Omega$ is small compared to the amplitude of the alternating voltage  ($\kappa eV_{\rm ac}^{\rm eff}/\Omega\gg 1$), multi-photon absorption occurs. Setting, $\kappa eV_{\rm ac}\,{\rm sin}\,\theta = n\Omega$, the sum over photon number can be approximated by an integral formula that corresponds to a classical detection of the RF field, \cite{Hamilton1970}
\begin{align*}
I(V)=\frac{1}{\pi}\int_{-\pi/2}^{\pi/2}\,I_{0}\left(V-|\Xi(\Omega)|V_{\rm ac}\,{\rm sin}\,\theta\,\right)\,{\rm d}\theta,
\end{align*}
where $I_0(V)$ is given by eq. (\ref{IV_eq}). This result offers a way to measure the magnitude of the Lehmann weight $\Xi(\Omega)$, where $|\Xi(\Omega)|$ is proportional to the sensitivity of the detector to RF power.\cite{Kanyolo2020JLPT} Conversely, this implies that our results are indispensable in dynamical Coulomb blockade experiments where long arrays are used as detectors of oscillating electromagnetic fields.\cite{Liou2014, LT28}

\section{Acknowledgement}
The authors wish to thank Prof. Yoshinao Mizugaki, Prof. Nobuhito Kokubo, Prof. Yuki Fuseya and Prof. Takeo Kato for insightful discussions, Bernard Kanyolo for help with simulations and Dr. Titus Masese and Edfluent services for proofreading the manuscript. This work was supported in part by JSPS KAKENHI Grant Number 18H05258, International Kyowa Scholarship Foundation, Ito Foundation for International Education Exchange, KDDI Foundation Scholarship for the International students and Monbukagakusho Honors Scholarship. 

\bibliography{CB}
\bibliographystyle{apsrev4-1} 


\begin{appendix}

\begin{widetext}

\section{\label{App: linear_response} Causal Linear Response}

Appendix \ref{App: linear_response} is meant for the skimming reader, who wants a quick reference to linear response (and its relevance to microwave power renormalization). Thus, we do not strive to introduce the entire subject of linear response and its subtleties. (For a comprehensive introduction to the subject, see ref. \citenum{Kubo1957}) 

Within Linear Response Theory, the response $\tilde{R}(t)$ of a system is related to the driving force, $\tilde{F}(t)$ by the central causal relation 
\begin{align}
\tilde{R}(t) = \int_{-\infty}^{t} \chi(t-s)\tilde{F}(s)ds, 
\label{LR_eq}
\end{align}
where $\chi(\tau)$ is the response function. The system variable, $\tilde{R}(t)$ obeys some equation of motion, 
\begin{align}
f(\partial/\partial t)\tilde{R}(t) = \tilde{F}(t),
\end{align}
with $f(\partial/\partial t)$ a function of $\partial/\partial t$. 
Introducing the Green's function of the system, $G_{\tilde{R}}(t)$, satisfying,
\begin{align}
f(\partial/\partial t) G_{\tilde{R}}(t-s) = \delta(t-s),    
\end{align}
we see that, 
\begin{multline}
f(\partial/\partial t)\tilde{R}(t) = \int_{-\infty}^{t}f(\partial/\partial t)G_{\tilde{R}}(t-s)\tilde{F}(s)ds 
= \int_{-\infty}^{t}\delta(t-s)\tilde{F}(s)ds =  \int_{0}^{+\infty}\delta(\tau)\tilde{F}(t - \tau)d\tau\\
= \int_{-\infty}^{+\infty}\delta(\tau)\theta(\tau + 0^+)\tilde{F}(t - \tau)d\tau = \theta(0^+)\tilde{F}(t) = \tilde{F}(t)    
\end{multline}
with $\theta(\tau)$ the Heaviside function. Thus, we can equate the Green's function to the response function: $\chi(t-s) = G_{\tilde{R}}(t - s)$. 

Substituting the Fourier transforms of $\tilde{R}(t) = \int d\omega \tilde{R}(\omega) \exp(-i\omega t)$ and $\tilde{F}(t) = \int d\omega \tilde{F}(\omega) \exp(-i\omega t)$ into eq. (\ref{LR_eq}), 
\begin{multline}
\int d\omega \tilde{R}(\omega) \exp(-i\omega t) = \int_{-\infty}^{t}\int \tilde{F}(\omega)\exp(-i\omega s) \chi(t-s) d\omega ds\\
\int \left \{ \tilde{F}(\omega) \int_{0}^{+\infty}\chi(\tau)\exp(i\omega\tau)d\tau \right \} \exp(-i\omega t) d\omega = \int \left \{ \tilde{F}(\omega)\,\Xi(\omega)\right \}\exp(-i\omega t) d\omega,
\end{multline}
where $\tau = t - s$ and,
\begin{subequations}
\begin{align}
\tilde{R}(\omega) = \Xi(\omega)\tilde{F}(\omega),\\
\Xi(\omega) = \int_{0}^{+\infty}\chi(\tau)\exp(i\omega\tau)d\tau = \int_{-\infty}^{+\infty}\theta(\tau)\chi(\tau)\exp(i\omega\tau)d\tau,\\
2\pi\theta(\tau)\chi(\tau) = \Xi(t).
\end{align}
\end{subequations}
Finally, that $\Xi(\omega) = \pm \exp(-\beta M)\exp i\eta(\omega)$ acts as the response function to the applied oscillating electromagnetic field is to be understood as the result of the arguments in section \ref{Vacuum Excitation}, and not necessarily the converse. This leaves the possibility that linear response is violated in complicated circuits, where novel physics may lurk.

\section{The Electromagnetic Environment in Large and Small Josephson Junctions}

Despite the existence of excellent reviews on the subject and techniques\cite{Vool2017, Ingold_Nazarov1992, Falci1991}, the authors found much of the techniques and prior concepts useful in following the arguments in this thesis scattered in various literature\cite{Zwanzig1973, Caldeira-Leggett1983, Green1954, Kubo1957, Zee2010, Leggett1984}. In particular, the techniques used in the subsequent chapters include path integral formalism\cite{ingold2002path} and Green's functions\cite{Leggett1984, Zee2010} to calculate phase-phase correlation functions and four-vector notation\cite{carroll1997} where Maxwell's equations appear for compactness. Thus, we include this section as a preamble for completeness and/or compactness. Hopefully it offers a more nuanced understanding of the Caldeira-Leggett model and $P(E)$ theory in the context of Green's functions and generally a path integral framework. 

\end{widetext}

\subsection{Organization}
Section \ref{large} considers how the effects of the electromagnetic environment arises via a normal current in large junctions. In the subsections, \ref{Josephson} introduces a 2-spinor and Pauli matrices that act on the spinor to derive the well-known Josephson equations. We proceed in \ref{Josephson-Maxwell} to introduce the effect of the environment as a normal current proportional to the electric field in the tunneling direction and a fluctuating noise current whose degrees of freedom we introduce in \ref{Environment} as a Caldeira-Leggett heat bath. 

Section \ref{small} tackles the environment in small junctions. In the subsections, \ref{Hamiltonian} introduces the total Hamiltonian of the Josephson junction including the environment and derives an expression for the tunneling current, \ref{Perturbation} expands this expression into a perturbation series and explicitly calculates the Cooper-pair kernel using the Pauli matrices introduced in \ref{Josephson} while \ref{path_integrals} uses path integral formalism to calculate phase-phase correlation functions in the $P(E)$ function. The expression for the Cooper-pair and quasi-particle tunneling current at finite temperature is derived in \ref{I_V}.

Note that, units where Planck's constant, Swihart velocity\cite{Swihart1961} and Boltzman constant are set to unity ($\hbar = \bar{c} = k_{\rm B} = 1$) and Einstein summation convention are used through out unless otherwise stated with ${\rm diag}\left \{ \eta_{\mu\nu} \right \} = (1, -1, -1, -1)$ the Minkowski space-time metric and $\eta_{\sigma\mu}\eta^{\sigma\nu} = \delta_{\mu}^{\nu}$ the Kronecker delta symbol.

\subsection{\label{large} Josephson Effect and the Electromagnetic Environment}

\subsubsection{\label{Josephson} The Josephson effect (large junction)} 

\begin{figure}[t!]
\begin{center}
\includegraphics[width=0.8\columnwidth,clip=true]{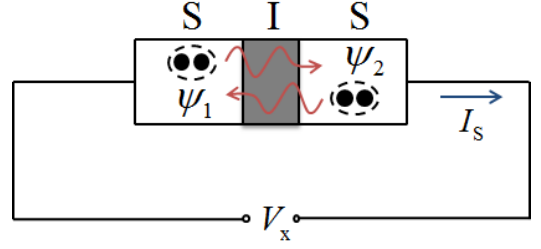}
  \caption{Schematic of a Josephson junction (S: superconductor, I: insulator, S: superconductor) depicting a Cooper pair from the left/right electrode tunneling through the insulator to the right/left electrode.}
\label{Schematic_JJ}
\end{center}
\end{figure}

The physics of Josephson junctions (schematic shown in Fig. \ref{Schematic_JJ}) is described by the well known Josephson equations\cite{Josephson1962},
\begin{subequations}\label{Josephson_eq}
\begin{align}
I_{\rm S} = 2eE_{\rm J}\sin 2\phi_{\rm x}(t)\\
\frac{\partial \phi_{\rm x}(t)}{\partial t} = eV_{\rm x}
\end{align}
\end{subequations}
Here, $2\phi_{\rm x}(t)$ denotes the phase difference across the junction, where the subscript $x$ distinguishes it from quantum phases of other circuit elements defined later in the manuscript, and $E_{\rm J}$ is the Josephson coupling energy\cite{Ambe_Bara1963}. The simplest derivation of eq.  (\ref{Josephson_eq}) follows from the real and imaginary parts of these two coupled Schrodinger equations 
\begin{subequations}\label{coupled_eq}
\begin{align}
i\frac{\partial \psi_1}{\partial t} = \mu_1\psi_1 + m_0 \psi_2,\\
i\frac{\partial \psi_2}{\partial t} = \mu_2\psi_2 + m_0 \psi_1.
\end{align}
\end{subequations}
Here, $\mu_1$ and $\mu_2$ and the chemical potentials of the left (1) and right (2) junction respectively, $\psi_1$ and $\psi_2$ are the Cooper-pair wavefunctions of the left and right superconductors respectively and $m_0$ is a coupling energy term characterizing magnitude of overlap for the two wavefunctions across the insulator. When a potential difference (voltage) $V_{\rm x}$ is applied across the junction, the two chemical potentials shift relative to each other in order to accomodate this change. This means that we can set $\mu_1 - \mu_2 = 2eV_{\rm x}$, where $2e$ is the Cooper pair charge. Based on this, it is instructive to define an average chemical potential, $\mu \equiv (\mu_1 + \mu_2)/2$, and solve for $\mu_1$ and $\mu_2$ in terms of $\mu$. This yields,
$\mu_1 = \mu + eV_{\rm x}$ and $\mu_2 = \mu - eV_{\rm x}$. Plugging this back to eq. (\ref{coupled_eq}) yields, 
\begin{subequations}\label{coupled_spinor_eq}
\begin{align}
i\frac{\partial \psi_1}{\partial t} = (\mu + eV_{\rm x})\psi_1 + m_0 \psi_2,\\
i\frac{\partial \psi_2}{\partial t} = (\mu - eV_{\rm x})\psi_2 + m_0 \psi_1.
\end{align}
\end{subequations}
From this, it is clear $\mu$ is simply the common chemical potential relative to which the voltage drop is measured. This observation implies we can set it to zero without loss of generality, $\mu = 0$.

The Cooper pair wavefunctions are defined as $\psi_1 = \sqrt{n_1}\exp(i2\varphi_1)$ and $\psi_2 = \sqrt{n_2}\exp(i2\varphi_2)$ where $n_1$ and $n_2$ are the number of Cooper-pairs in the left (1) and right (2) superconductor respectively, and $2\varphi_1$ and $2\varphi_2$ is their respective macroscopic quantum phases. Plugging these definitions into eq. (\ref{coupled_spinor_eq}), we find,
\begin{subequations}
\begin{align}
\frac{\partial n_1}{\partial t} = 2m_0\sqrt{n_1n_2}\sin(2[\varphi_2 - \varphi_1])\\
\frac{\partial n_2}{\partial t} = 2m_0\sqrt{n_2n_1}\sin(2[\varphi_1 - \varphi_2])
\end{align}
\end{subequations}
and,
\begin{subequations}
\begin{align}
2\frac{\partial \varphi_1}{\partial t} = -m_0\sqrt{\frac{n_2}{n_1}}\cos(2[\varphi_2 - \varphi_1]) - eV_{\rm x},\\
2\frac{\partial\varphi_2}{\partial t} = -m_0\sqrt{\frac{n_2}{n_1}}\cos(2[\varphi_2 - \varphi_1]) + eV_{\rm x}.
\end{align}
\end{subequations}
Taking the approximation that any tunneling currents that arise have the effect of varying $n_1$ and $n_2$ by only a small amount $(\delta n)^2 \simeq 0$ from an equilibrium value given by $\frac{1}{2}(n_1 + n_2) \equiv n$, i.e. $n_1 = n + \delta n$ and $n_2 = n - \delta n$, we see that the supercurrent across the barrier is given by $I_{\rm S} \equiv e\partial \delta n/\partial t$ and the voltage drop by $\partial(\varphi_2 - \varphi_1) /\partial t = eV_{\rm x}$, which yield eq. (\ref{Josephson_eq}) when $E_{\rm J} = nm_0$ and $\varphi_2 - \varphi_1 = \phi_{\rm x}$. 

There is another advantage of setting $\mu = 0$. In particular, eq. (\ref{coupled_spinor_eq}) becomes a spinor equation, 
\begin{subequations}\label{Schrodinger_eq}
\begin{align}
i\frac{\partial \psi}{\partial t} = H_{\rm cp} \psi\\
H_{\rm cp} = eV_{\rm x} \sigma_3 + m_0 \sigma_1, \\
\psi \equiv \begin{pmatrix}
\psi_1 \\ 
\psi_2
\end{pmatrix},
\label{J_Hamiltonian_eq}
\end{align}
\end{subequations}
where $\sigma_1$ and $\sigma_3$ are the Pauli matrices,
\begin{subequations}\label{Pauli_eq}
\begin{align}
\sigma_1 = \begin{pmatrix}
0 & 1\\ 
1 & 0
\end{pmatrix},\; \; 
\sigma_3 = \begin{pmatrix}
1 & 0\\ 
0 & -1
\end{pmatrix}
.
\end{align}
Since $E_{\rm J} \propto m_0$, $\sigma_{1}$ plays the role of the coupling term that results to Cooper-pair tunneling. Using the last Pauli matrix, 
\begin{align}
\sigma_2 = \begin{pmatrix}
0 & -i\\ 
i & 0
\end{pmatrix},\; \;
\end{align}
\end{subequations}
we can define two operators $\sigma_+ \equiv (\sigma_1 + i\sigma_2)/2$ and $\sigma_- \equiv (\sigma_1 - i\sigma_2)/2$. These operators form tunneling matrix elements with the spinor and a transpose conjugate spinor defined as,
\begin{align}
\psi^{\dagger} \equiv \psi^{*\rm T} = (\psi_1^*, \psi_2^*). 
\end{align}
For instance, tunneling from left to right requires replacing $\psi_1$ with $\psi_2$ and annihilating $\psi_1$, which corresponds to $\sigma_+\psi$. The inverse process process corresponds to $\sigma_-\psi$. These matrices will be useful when calculating Cooper-pair tunneling rates for small junctions (See eq. \ref{alpha2_eq}).

\subsubsection{\label{Josephson-Maxwell} Sources of the Electromagnetic Field as the Josephson Junction Environment} 

Equation \ref{Josephson_eq} only considers the superconducting current and thus neglects the environment that lead to effects such as Coulomb blockade. The environment consists of all sources of the electromagnetic field (including the field itself) which couple to the Cooper-pair wavefunction via the phase difference thus determining the $I-V$ characteristics satisfying eq.  (\ref{Josephson_eq}). Specifically, the environment arises from processes such as the alternating currents and voltages, thermal fluctuations in the form of Johnson-Nyquist noise and coupled high impedance circuit environments\cite{Nyquist1928, Koval2010}. 

Using eq.  (\ref{Josephson_eq}), one can define a conserved energy by treating the junction as a capacitance  
\begin{subequations}\label{Josephson_eq Hamiltonian}
\begin{align}
\label{J_Hamiltonian2_eq}
E = Q_{\rm x}^2/2C - E_{\rm J}\cos(2\phi_{\rm x})\\
Q_{\rm x} = -CV_{\rm x}\\
\frac{\partial Q_{\rm x}}{\partial t} = I_{\rm S}
\end{align}
\end{subequations}
where $C$ is the capacitance of the junction. 
Modifying the last equation in (\ref{Josephson_eq Hamiltonian}) to
\begin{equation}
\frac{\partial Q_{\rm x}}{\partial t} = \sum_{\rm a} I_{\rm a},
\label{current_sum_eq}
\end{equation}
one can then include all the environmental sources of energy in the form of currents. In fact, to arrive at eq.  (\ref{current_sum_eq}), the phase-difference needs to couple to the electromagnetic field (Maxwell equations) in a straight-forward manner
\begin{subequations}\label{Josephson_Maxwell_eq}
\begin{align}
\frac{\partial \phi_{\rm x}}{\partial x^{\mu}} = ed_{\rm eff}N^{\nu}F_{\mu \nu}\\
N^{\mu}N_{\mu} = -\sum_{ij} N_{i}N_{j}\delta_{ij} = -1 \\
\frac{\partial F^{\mu \nu}}{\partial x^{\mu}} = -\frac{1}{\varepsilon_0 \varepsilon_{\rm r}} \sum_{\rm a} J_{\rm a}^{\nu}
\end{align}
\end{subequations}
Here, $F_{\mu \nu} = \partial A_{\mu}/\partial x_{\nu} - \partial A_{\nu}/ \partial x_{\mu} = -F_{\nu \mu}$ is the electromagnetic tensor, $d_{\rm eff}$ is the thickness of the barrier 
and $N^{\mu} = (0, N^i)$ points in the direction $N^i$ normal to the tunnel barrier. We have used Einstein notation where only the Greek indices are summed over and the Minkowski space-time signature is ${\rm diag}(\eta_{\mu \nu})$ = (+, -, -, -).
Taking the total derivative $\eta^{\mu\nu}\partial/\partial x^{\nu} = \partial/\partial x_{\nu}$ of eq.  (\ref{Josephson_Maxwell_eq}a) (with $\eta_{\mu\alpha}\eta^{\nu\alpha} = \delta_{\mu}^{\nu}$) and using $\partial N^{\rm \nu}/\partial x^{\mu} = 0$ , we arrive at \begin{equation}\label{EOM_eq}
\eta^{\mu\nu}\frac{\partial^2 \phi_{\rm x}}{\partial x^{\mu} \partial x^{\nu}} = \frac{\partial^2 \phi_{\rm x}}{\partial x^{\mu} \partial x_{\mu}} = -\frac{ed_{\rm eff}}{\varepsilon_0 \varepsilon_{\rm r}} \sum_{\rm a} N_{\mu}J_{\rm a}^{\mu} = -wJ
\end{equation}   
which is eq.  (\ref{current_sum_eq}) in disguise. Note that $F_{0 i} = \vec{E}$ and $\frac{1}{2}\sum_{i,j}\varepsilon_{i j k} F^{i j} = \vec{B}$ where $\vec{E}$ and $\vec{B}$ are the $x, y, z$ components of the electric and magnetic fields respectively and $\varepsilon_{ijk}$ is the Levi-Civita symbol. eq.  (\ref{EOM_eq}) is the sourced Klein-Gordon equation with $J = \sum_{\rm a} N_{\mu}J_{\rm a}^{\mu}$ the source and $w = ed_{\rm eff}/\varepsilon_0 \varepsilon_{\rm r}$ the coupling constant. The vector $N_{\mu}$ and the anti-symmetry of the electromagnetic tensor $F_{\mu \nu}$ guarantees that, unlike Maxwell equations, the coupled Klein-Gordon equation lives in 2+1 dimensions instead of 3+1. For instance, when the tunnel barrier is aligned to the y-z direction, $N^{\mu} = (0, 1, 0, 0)$ and eq.  (\ref{Josephson_Maxwell_eq}a) and by extension eq.  (\ref{EOM_eq}) become independent of $x$.
\begin{equation}
\frac{\partial^2 \phi_{\rm x}}{\partial x^{\mu} \partial x_{\mu}} = 
\frac{\partial^2 \phi_{\rm x}}{\partial t^2}-\frac{\partial^2 \phi_{\rm x}}{\partial y^2}-\frac{\partial^2 \phi_{\rm x}}{\partial z^2}
= -wJ    
\end{equation}
Furthermore, taking the limit for small junctions which corresponds to taking the area of the barrier $\mathcal{A}$ to be small such that the phase neither varies with $y$ nor $z$, we arrive at eq.  (\ref{current_sum_eq}) \begin{subequations}
\begin{align}
\frac{\partial^2 \phi_{\rm x}}{\partial t^2}= -4E_{\rm c}E_{\rm J} \sin (2\phi_{\rm x})-\frac{1}{RC}\frac{\partial \phi_{\rm x}}{\partial t} - 2E_{\rm c}I_{\rm F}/e \\
\sum_{\rm a}^3 J_{\rm a}^{1} = J_{\rm S}^{1}+ J_{\rm N}^{1} + J_{\rm F}^{1}\\
\left \langle I_{\rm F}(t)I_{\rm F}(t') \right \rangle = \frac{4 \beta^{-1}}{R}\delta(t-t')
\end{align}
\label{RCSJ_eq}
\end{subequations}
with $J_{\rm S}^{1} = J_{\rm cp}\sin2\phi_{\rm x}$ the supercurrent, $J_{\rm N}^{1} = \sigma_{xx} F_{0 1}$ the normal current and $\sigma_{xx}$ the effective conductivity of the barrier along the $x$ direction. Here, we have used the cross-sectional area of the junction, $\mathcal{A}$ to define $J^{\mu}_{\rm a}\mathcal{A} = I^{\mu}_{\rm a}$, the junction capacitance $C = \varepsilon_0 \varepsilon_{\rm r}\mathcal{A}/d_{\rm eff}$, the charging energy $E_{\rm c} = e^{2}/2C$ and the junction conductance $1/R =  \sigma_{xx} \mathcal{A}/d_{\rm eff}$. Finally, $\beta^{-1} = k_{\rm B}T$ is the inverse temperature and we have assumed the fluctuation current $J_{\rm F}^1\mathcal{A} = I_{\rm F}$ is Gaussian-correlated over a bath (B stands for bath or Boltzmann distribution) with the thermal correlation function given by eq.  (\ref{RCSJ_eq}c).

\subsection{\label{Environment} Generalized Impedance Environment and the Thermal Bath
}

It is straight forward to generalize the conductance $1/R$ in eq.  (\ref{RCSJ_eq}) using a spectral function $K(\omega) = [Z^{-1}(\omega) + Z^{-1}(-\omega)]/(2\pi)$ describing the macroscopic physics of the microscopic degrees of freedom of the system undergoing Brownian motion due to a heat bath comprising $k$ harmonic oscillators,\cite{Zwanzig1973}
\begin{subequations}\label{Heat_bath_eq}
\begin{align}
H_{\rm B} = \sum_{n=1}^{k}\left \{ \frac{Q_{n}^2}{2C_{n}}+\frac{(\phi_{n}-\phi_{\rm x})^2}{2e^2L_{n}} \right \} \\
K(t) =\sum_{n=1}^{k}L_{n}^{-1} \cos (\omega_{n}t)=\int_{-\infty}^{+\infty}d\omega K(\omega) \exp-i\omega t 
\end{align}
\end{subequations}
where $H_{\rm B}$ is the Hamiltonian of the heat bath consisting of $L_nC_n$ circuits in parallel where $\omega_n = 1/L_nC_n$, $Q_{n}$, $\phi_n$ are the charges stored by and the phases of the elements and $K(t)$ is referred to as the Kernel representing the dissipative nature of the circuit. The generalized Lagrangian for the system is given by
\begin{multline}\label{g_lagrangian_eq}
\mathcal{L}=\frac{C}{2e^2}\left (\frac{\partial \phi_{\rm x}(t)}{\partial t} \right )^2
 -\frac{1}{2e^2}\int_{-\infty}^{+\infty}\phi_{\rm x}(t)\frac{\partial K(s-t)}{\partial s}\phi_{\rm x}(s)ds\\
 -\frac{1}{e}\int_{-\infty}^{+\infty}I_{\rm F}(t - s)\phi_{\rm x}(s)ds + E_{\rm J}\cos(2\phi_{\rm x}), 
\end{multline}
where the fluctuation current is given by, 
\begin{subequations}
\begin{align*}
I_{\rm F}(t) = \sum_{n=1}^{k} \left \{\omega_n Q_n \sin(\omega_nt) + e^{-1}L_n^{-1} (\phi_n - \phi_{\rm x})\cos(\omega_nt) \right \}\\
\left \langle I_{\rm F}(t)I_{\rm F}(t') \right \rangle = \sum_{n=1}^{k}2L_{n}^{-1}
\left \langle H_{\rm B}(\omega_{n}) \right \rangle\cos\omega_{n}(t-t').
\end{align*}
The average is over the thermal bath degrees of freedom. For the Ohmic conductance above, we have $Z^{-1}(\omega) = 1/R$ and $\left \langle H_{\rm B}(\omega) \right \rangle = \beta^{-1}$ where the continuous, large $k$ limit
\begin{equation}\label{large N_limit_eq}
\lim_{k \to \infty} \sum_{n=1}^{k} L_{n}^{-1}\times \rightarrow \int_{-\infty}^{\infty}d\omega K(\omega) \times
\end{equation}
is taken in accordance with eq.  (\ref{Heat_bath_eq}b) thus recovering eq.  (\ref{RCSJ_eq}c).
The fluctuation current density certainly satisfies the Green-Kubo relation\cite{Green1954, Kubo1957},
\begin{multline}
\sigma_{xx} = \frac{\beta}{4} \int d\,^4xN^{\nu}N^{\mu}\left \langle J_{{\rm F}\nu}(t)J_{{\rm F}\mu}(0) \right \rangle\\
= \frac{\beta}{4}\mathcal{A}^{-1}d_{\rm eff}\int dt \left \langle I_{\rm F}(t)I_{\rm F}(t') \right \rangle = \frac{d_{\rm eff}}{R\mathcal{A}},
\label{Green_Kubo_eq}
\end{multline}
\end{subequations}
where we have used eq.  (\ref{RCSJ_eq}c) in the last line.
Note that to obtain the correct equation of motion, integration by parts of the second term in eq.  (\ref{g_lagrangian_eq}) should be performed after applying the Euler-Lagrange equations, then the boundary term is dropped
\begin{equation}\label{b_term_eq}
\frac{1}{e^2}\int_{-\infty}^{+\infty}\frac{\partial }{\partial s}\left [ K(s-t)\phi_{\rm x}(s) \right ]ds = 0 
\end{equation}

\subsection{\label{small} Coulomb Blockade and the Electromagnetic Environment}

\subsubsection{\label{Hamiltonian} Hamiltonian}

Consider a mesoscopic tunnel junction with capacitance $C$ driven by a voltage source $V_{\rm x}$ via an environmental impedance $Z(\omega)$. Each circuit element is characterized by a phase $\phi_{\rm a}$ related to the voltage drop $V_{\rm a}$ of the element in the circuit by $\phi_{\rm a}(t)=\int_{-\infty}^{t} eV_{\rm a}(\tau) d\tau$, where the subscript a = J, x or z corresponds to the junction, voltage source and environment impedance and $\kappa e = 2e, e$ corresponds to Cooper pair, quasi-particle charge respectively. That the effect of the environmental impedance $Z(\omega)$ can be represented by a single quantum phase $\phi_{\rm z}$ defined by the voltage drop over $Z(\omega)$ is not at all obvious. At this stage, we treat it as an ansatz. It will not appear in the equations until we impose the topological constraint $\sum_a \phi_a = e\Phi$ on the circuit. The voltages $V_{\rm z}$ and $V_{\rm J}$ decrease as one moves clockwise along the circuit, whereas the value increases for the voltage source $V_{\rm x}$ in the same direction. The corresponding charge on the junction is defined as $Q_{\rm J} = CV_{\rm J}$ where $C$ is the capacitance of the junction. The circuit can store a topological flux $e\Phi = \phi_{\rm J} + \phi_{\rm z} + \phi_{\rm x} = \sum_{a}\phi_{\rm a}$ related to a topological potential $\int_{-\infty}^{t} A(\tau)d\tau = \Phi(t)$, which leads to the renormalization effect in section \ref{Bloch_section}. 

The total Hamiltonian, $\mathcal{H}$ of the circuit (Fig. \ref{Caldeira_fig}) is given by the expression,
\begin{equation}
\mathcal{H} = \sum_{\kappa = 1}^{2} H_{\kappa}  + H_{\rm J} + H_{\rm z} 
\label{H_eq}
\end{equation}
Here, $\sum_{\kappa = 1}^{2} H_{\kappa} = H_1 + H_2 $ where the Cooper-pair Hamiltonian $
H_2 = \mu \psi^{+}\psi = \psi^{+}i\left [\partial/\partial t+iH_{\rm cp}\right ]\psi$ depends on the chemical potential $\mu$ and the 2-spinor $\psi$ 
and the quasi-particle Hamiltonian $H_1$ is given by,
\begin{align}
H_1  = H_{\rm L} + H_{\rm R} 
=\sum_{p,\sigma}\epsilon_{p\sigma} \gamma_{p\sigma}^\dagger \gamma_{p\sigma}+\sum_{q,\sigma}\epsilon_{q\sigma} \gamma_{q\sigma}^\dagger \gamma_{q\sigma}
\label{sum_H_1}   
\end{align}
where $\gamma_{p\sigma}$ or $\gamma_{q\sigma}$ and $\gamma_{p\sigma}^\dagger$ or $\gamma_{q\sigma}^\dagger$ are the annihilation and creation operators respectively of a quasi-particle state with energy $\epsilon_{p\sigma}$ or $\epsilon_{q\sigma}$, momentum $p$ or $q$ and spin $\sigma$ in the left or right electrode,   
\begin{equation}
H_{\rm J} = \sum_{\kappa = 1}^{2}\Theta_\kappa\exp(-i\kappa \phi_{\rm J}) + h.c.
\label{H_J_eq}
\end{equation}
is the tunneling Hamiltonian where 
\begin{subequations}
\begin{equation}
\Theta_1 = \sum_{p \neq q,\sigma} M_{pq}
\gamma_{q\sigma}\gamma_{p\sigma}^\dagger 
\label{operator1_eq}
\end{equation} 
\begin{equation}
\Theta_2 = \frac{E_{\rm J}}{2}(\sigma_1-i\sigma_2) \equiv \frac{E_{\rm J}}{2}\sigma_-
\label{operator2_eq}
\end{equation}
\end{subequations}
$\sigma_{1,2}$ are the $x, y$ Pauli matrices acting on the 2-spinor given in eq. (\ref{Pauli_eq}), $M_{pq}$ is a dimensionful spin-conserving complex-valued quasi-particle tunneling matrix, $p \neq q$ enforces the condition $[H_{\rm L}, H_{\rm R}] = 0$ and $E_{\rm J}$ is the Josephson coupling energy\cite{Ambe_Bara1963}, 
\begin{multline}
H_{\rm z}=\frac{(Q_{\rm J} + CV_{\rm x} + CA)^2}{2C}\\
+ \sum_{n=1} \left\{\frac{Q_n^2}{2C_n} + e^{-2}\frac{(\phi_n - \phi_{\rm J} + \phi_{\rm x} + e\Phi)^2}{2L_n}\right\}
\label{H_Z_eq}
\end{multline}
is the Hamiltonian describing the environmental impedance $Z(\omega)$ and junction capacitance $C$, where $Z(\omega)$ is characterized by an infinite number of parallel $L_nC_n$ circuits coupled serially to the tunnel junction. One can define $Q = Q_{\rm J} - CV_{\rm x}$ and $\phi = \phi_{\rm J} - \phi_{\rm x}$ as the fluctuation variables of the junction charge $Q_{\rm J} = CV_{\rm J}$ and junction phase $\phi_{\rm J}(t) = e\int_{-\infty}^{t}dt'V_{\rm J}(t')$ around the mean value determined by the voltage source $V_{\rm x}$, where $V_{\rm J}(t')$ is the voltage drop across the junction. (See ref. \citenum{Ingold_Nazarov1992} on page 27. Note that $\phi_{\rm J}$ and $\phi$ are related by a suitable unitary transformation $\mathcal{U}$ of the Hamiltonian,
\begin{equation}
\mathcal{H'} = i\mathcal{U^{\dagger}}\frac{\partial }{\partial t}\mathcal{U}+\mathcal{U}\mathcal{H}\mathcal{U^{\dagger}},
\label{unitarity_eq}
\end{equation}
where $\mathcal{H'} = H'_{1}+ H_{2} + H_{\rm J} + H_{\rm z}$, $H_{\rm J} = \sum_{\kappa = 1}^{2}\Theta_\kappa\exp(-i\kappa \phi_{\rm J}) + h.c.$, $H'_{1} = \sum_{p \neq q, \sigma}\epsilon'_{p\sigma} \gamma_{p\sigma}^\dagger \gamma_{p\sigma}+\sum_{p \neq q,\sigma}\epsilon_{q\sigma} \gamma_{q\sigma}^\dagger \gamma_{q\sigma}$ and $\epsilon'_{p\sigma} = \epsilon_{p\sigma} + eV_{\rm x}$. $Q = Q_{\rm J} - CV_{\rm x}$, $Q_n$ are the conjugate variables to $\phi = \phi_{\rm J} - \phi_{\rm x}$, $\phi_{\rm n}$ satisfying the charge-phase commutation relation,
\begin{align}\label{commutation_eq}
[\phi_{n},Q_{m}]=i\delta_{mn} e,\\
[\phi, Q]= ie
\end{align}
where $\delta_{ab}$ is the Kroneker delta. Operators, $O(t)$ in the Heisenberg picture are related to the ones in the Schr\"{o}dinger picture, $O(0)$ by $O(t)=U_0(t)^\dagger O(0)U_0(t)$ with the unitary evolution operator $U_0(t)$ given by $U_0(t)=\exp\left\{-i\sum_{\kappa = 1}^{2} H_{\kappa} t\right\}$ in the absence of tunneling. 

In what follows, we assume the Cooper-pair ground state energy $\mu = 0$, as we did in eq. \ref{coupled_spinor_eq}. The tunneling current $I(V)$ at the junction is given by 
\begin{equation}
I(V, s) = {\rm tr}\left \langle  \mathcal{T}\left \{U^{\dagger}I_{\rm J}(0)U\right \} \right \rangle   
\label{current_eq}
\end{equation}
Here, $U = U_0 + U_{\rm int}$ where
\begin{widetext}
\begin{multline}
U_{\rm int} = \exp\left (-i\int_{-s}^{+s}dt\, H_{\rm J}(t)\right) 
= \exp\left (-i\int_{-s}^{0}dt\, H_{\rm J}(t) \right)\exp\left (-i\int_{0}^{s}dt\, H_{\rm J}(t) \right) \\
= \exp\left (+i\int_{0}^{-s} H_{\rm J}(t)dt\,\right)\exp\left (-i\int_{0}^{s}dt\, H_{\rm J}(t)\right) 
= U^{\dagger}_{\rm int}(-s)U_{\rm int}(+s)    
\end{multline}
\end{widetext}
$\mathcal{T}$ is the time ordering operator with the property given by $\mathcal{T} \Theta_{\kappa}(t) \Theta_{\kappa}^{\dagger}(0) = \Theta_{\kappa}(t) \Theta_{\kappa}^{\dagger}(0)$,
$\mathcal{T} \Theta_{\kappa}(0) \Theta_{\kappa}^{\dagger}(t) = \Theta_{\kappa}^{\dagger}(t) \Theta_{\kappa}(0)$. Here, $s$ is the elapsed time after switching on the interaction term $U_{\rm int}(s)$ and takes the range $0 \leq s \leq +\infty$. [Note that $U^{\dagger}_{\rm int}(s)$ takes care of $c.c.$ term in eq. (\ref{current_perturb_eq}), updating the integral range as discussed: $\int_{0}^{s}dt \rightarrow \int_{-s}^{0}dt + \int_{0}^{s}dt = \int_{-s}^{+s}dt$.] We shall be interested in the  current $I(V, s \rightarrow +\infty) = I(V)$ at equilibrium [eq. (\ref{IV_eq})].

The tunneling current operator is 
\begin{equation}
I_{\rm J}(0) = -i[Q_{\rm J}(0), H_{\rm J}(0)],
\label{current_Zero_eq}
\end{equation}
and the average $\langle... \rangle$ is over, the quasi-particle equilibrium states, whose density matrix is given by $\rho_1 = \rho_{\rm L}\rho_{\rm R} = \mathcal{Z}_1^{-1}\exp(-\beta H_1)$ with $ \mathcal{Z}_1 = \mathcal{Z}_{\rm L}\times \mathcal{Z}_{\rm R} = \prod_{p} [1 + \exp(-\beta \epsilon_{ps})]\times \prod_{q}[1 + \exp(-\beta \epsilon_{qs})]$, and the environment $\rho_{\rm env} = \mathcal{Z}_{\rm env}^{-1}\exp(-\beta H_{\rm z})$ where $\beta=1/k_{\rm B}T$ is the inverse temperature while the trace (tr) is over the Pauli matrices.

\subsubsection{\label{Perturbation} Perturbation Expansion}

We can then expand eq. (\ref{current_eq}) as a perturbation series in the tunneling Hamiltonian $H_{\rm J}(t)$, 
\begin{equation}
I = \left \langle I_{\rm J}(0) -i \int_{-\infty}^{+\infty} [I_{\rm J}(0), H_{\rm J}(t)]dt + O(H_{\rm J}^2) \right \rangle.
\label{current_perturb_eq}
\end{equation}
Using $\mathcal{T} \Theta_{\kappa}(t) \Theta_{\kappa}^{\dagger}(0) = \Theta_{\kappa}(t) \Theta_{\kappa}^{\dagger}(0)$,
$\mathcal{T} \Theta_{\kappa}(0) \Theta_{\kappa}^{\dagger}(t) = \Theta_{\kappa}^{\dagger}(t) \Theta_{\kappa}(0)$,
$\left \langle \Theta_{\kappa}(t) \right \rangle = 0$ and  
$\left \langle \Theta_{\kappa'}(t) \Theta_{\kappa}^{\dagger}(0) \right \rangle = \left \langle \Theta_{\kappa'}^{\dagger}(t) \Theta_{\kappa}(0) \right \rangle = \alpha_{\rm \kappa} (t)\delta_{\kappa' \kappa}$,
we find that 
\begin{multline}
I \simeq \left \langle \int_{-\infty}^{+\infty} dt\, [H_{\rm J}(t),[Q_{\rm J}(0),H_{\rm J}(0)]] \right \rangle\\
= ie\sum_{\kappa = 1}^{2}  \int_{-\infty}^{+\infty}dt\,\biggl(\alpha_{\kappa} (t)\left \langle \sin \left [ \kappa \Delta\phi_{\rm J}(t) \right ] \right \rangle_{\phi_{\rm J}} \biggr) 
\label{calculation_eq}
\end{multline}
where $\Delta \phi_{\rm J}(t) = \phi_{\rm J}(t)-\phi_{\rm J}(0)$. Thus, the `particle' degrees of freedom $\alpha_{\kappa}(t)$ and the environment are decoupled and the trace over the environment $\left \langle...\: \rho_{\rm env}\right \rangle$ has been re-written as $\left \langle...\right \rangle_{\phi_{\rm J}}$ in terms of the junction phase $\phi_{\rm J}$ degree of freedom.
(section \ref{path_integrals})

For the quasi-particle current, the kernel $\alpha_1(t)$ scales with the dimensionless tunneling conductance $e^{-2} R_{\rm T}^{-1}$ but its functional form depends on the gap, reflecting the corresponding structures in the quasi-particle $I-V$ characteristics. It can be computed by taking the continuous limit,
$2\pi e^2R_{\rm T}\mathcal{N}_{\rm L}(0)\mathcal{N}_{\rm R}(0)M_{pq}M_{qp}^* \rightarrow 1$,
\begin{widetext}
\begin{multline}
\alpha_1(t) = \left \langle \Theta_{1}(t) \Theta_{1}^{\dagger}(0) \right \rangle
= \sum_{p \neq q, \sigma}\sum_{p' \neq q', \sigma}M_{pq} M_{q'p'}^{*}
\langle R,s|\langle L,s|\gamma_{q\sigma}(t)\gamma_{p\sigma}^{\dagger}(t)\gamma_{p'\sigma}(0)\gamma_{q'\sigma}^{\dagger}(0)\rho_1|L,s\rangle|R,s\rangle \\
= 2\sum_{p \neq q}\sum_{p' \neq q'} M_{pq} M_{q'p'}^{*}
\langle R|\gamma_{q}(t)\langle L|\gamma_{p}^{\dagger}(t)\gamma_{p'}(0)\rho_{\rm L}|L \rangle\gamma_{q'}^{\dagger}(0)\rho_{\rm R}|R \rangle 
= 2\sum_{p \neq q}f(\epsilon_{p})\exp\left \{-i\epsilon_{p} t\right \}
\sum_{p' \neq q'} M_{pq} M_{q'p'}^{*} \delta_{pp'}
\langle R|\gamma_{q}(t)\gamma_{q'}^{\dagger}(0)\rho_{\rm R}|R \rangle \\
= 2\sum_{p \neq q}f(\epsilon_{p})[1-f(\epsilon_{q})]\exp\left \{i(\epsilon_{q}-\epsilon_{p})t\right \} \sum_{p' \neq q'} M_{pq} M_{q'p'}^{*} \delta_{qq'} \delta_{pp'}
= 2\sum_{p \neq q}f(\epsilon_{p})[1-f(\epsilon_{q})]M_{pq} M_{qp}^{*}\exp\left \{i(\epsilon_{q}-\epsilon_{p})t\right \}\\
\rightarrow \frac {1}{\pi e^2R_{\rm T}}\int_{-\infty}^{+\infty}\int_{-\infty}^{+\infty}d\epsilon_{p}d\epsilon_{q}\frac{\mathcal{N}_{\rm L}(\Delta)}{\mathcal{N}_{\rm L}(0)}\frac{\mathcal{N}_{\rm R}(\Delta)}{\mathcal{N}_{\rm R}(0)}
f(\epsilon_{p})(1-f(\epsilon_{q}))\exp\left \{i(\epsilon_{q}-\epsilon_{p})t\right \}.
\label{alpha_eq}
\end{multline}
\end{widetext}
Here, $e^2R_{\rm T}$ is the dimensionless tunnel resistance, $f(E) = [1+\exp(\beta E)]^{-1}$ is the Fermi-Dirac function and $\mathcal{N}_{\rm L}(\Delta), \mathcal{N}_{\rm R}(\Delta)$ is the left, right BCS density of states\cite{BCS1957} which reduce to the electron density of states $\mathcal{N}_{\rm L}(0), \mathcal{N}_{\rm R}(0)$ when the superconducting gap $\Delta = 0$ vanishes, 
\begin{subequations}\label{Jacobian_eq}
\begin{align}
\frac{dE_{p}}{d\epsilon_{p}}\frac{dE_{q}}{d\epsilon_{q}} = \frac{\mathcal{N}_{\rm L}(\Delta)}{\mathcal{N}_{\rm L}(0)}\frac{\mathcal{N}_{\rm R}(\Delta)}{\mathcal{N}_{\rm R}(0)}\\
E_{p} = \sqrt{\epsilon_{p}^2-\Delta^2}, E_{q} = \sqrt{\epsilon_{q}^2-\Delta^2}
\end{align}
\end{subequations}
where $E_{p} = p^2/2m, E_{q} = q^2/2m$ is the kinetic energy of the electrons above the Fermi sea. 

Likewise, calculating $\alpha_2(t)$, we find,
\begin{multline}\label{alpha2_eq}
\alpha_{\rm 2} = \left \langle \Theta_2^{\dagger}(t)\Theta_2(0) \right \rangle = \left \langle \Theta_2^{\dagger}(0)\Theta_2(0) \right \rangle \\
= (\frac{E_{\rm J}}{2})^2{\rm tr}\left \{(\sigma_1 + i\sigma_2)(\sigma_1 - i\sigma_2) \right \}\\
= \frac{E_{\rm J}^2}{4}{\rm tr}\left \{2\sigma_0+i[\sigma_2,\sigma_1]\right \}
= \frac{E_{\rm J}^2}{2}{\rm tr}\left \{\sigma_0+\sigma_3\right \} = E_{\rm J}^2.
\end{multline}
We discover that, unlike $\alpha_1(t)$, $\alpha_2(t) = \alpha_2(0)= E_{\rm J}^2$ is time independent and only depends on the strength of Cooper pair tunneling, $E_{\rm J}$.

\subsubsection{\label{path_integrals} Path Integrals and Phase Correlations}

To calculate the remaining average over $\phi_{\rm J}$ in eq. (\ref{calculation_eq}), we work in Minkowski time at zero temperature (thus by-passing a rigorous but otherwise tedious Wick rotation to Euclidean time) since the finite temperature propagator is trivially related to the zero temperature result [eq. (\ref{propagator_finite_eq}) for the trivial relation and Appendix \ref{App:Functional_integral} for the formalism].

In this formalism, given an observable $O(\phi_{\rm J})$, its average at zero temperature is given by the functional/path integral,
\begin{equation}
\lim_{\beta\rightarrow \infty}\langle O\rangle_{\phi_{\rm J}}
= \mathcal{Z}^{-1}\prod_{n=1}^{k}\int D\phi_{\rm n} D\phi_{\rm J}O(\phi_{\rm J})\exp iS_{\rm z}(\phi_{n}, \phi_{\rm J}),
\label{avrg_eq}
\end{equation}
where $\mathcal{Z}=\prod_{n=1}^{k}\int D \phi_{\rm n}D\phi_{\rm J}\exp iS_{\rm z}(\phi_{\rm n}, \phi_{\rm J})$ is the partition function normalizing eq. (\ref{avrg_eq}) and the Lagrangian in the action for the environment $S_{\rm z}(\phi_{\rm n}, \phi_{\rm J})$ is given by the (inverse) Legendre transform of the environment Hamiltonian in eq. (\ref{H_Z_eq})
\begin{subequations}
\begin{align}\label{Legendre_eq}
S_{\rm z} = \int \mathcal{L}_{\rm z} dt = \int \left \{(Q_{\rm J} + Q_{\rm x} - CA)\frac{\partial H_{\rm z}}{\partial Q_{\rm J}} - H_{\rm z}\right \}dt,\\
C\frac{\partial \phi_{\rm x}(t)}{\partial t} = eQ_{\rm x}, C\frac{\partial \phi_{\rm J}(t)}{\partial t} = eQ_{\rm J}, \frac{\partial \Phi(t)}{\partial t} = A(t).
\end{align}
\end{subequations}
The effective action $S'_{\rm z}(\phi_{\rm J})$ resulting from performing first the functional integral product over $\phi_n$ is given by
\begin{subequations}
\begin{widetext}
\begin{multline}\label{S_circuit_eq}
S'_{\rm z}(\phi - e\Phi)
= S_{\rm z}^0(\phi-e\Phi) + S_{\rm z}^{\rm int}(\phi-e\Phi)
= \frac{C}{2e^2}\int_{-\infty}^{+\infty}\left(\frac{\partial [\phi(t)-e\Phi(t)]}{\partial t}\right)^2dt 
- \frac{1}{2e^2}\int_{-\infty}^{+\infty}\frac{\left [ \phi(t)-e\Phi(t) \right ]^2}{\sum_n L_n}dt\\
-\frac{1}{4\pi e^2}\int_{-\infty}^{+\infty}\int_{-\infty}^{+\infty}[\phi(t) - e\Phi(t)]\frac{\partial Z^{-1}(s-t)}{\partial s}[\phi(t)-e\Phi(t)]dsdt
+ \frac{1}{e}\int_{-\infty}^{+\infty}I_{\rm F}(t)[\phi(t)-e\Phi(t)]dt
\end{multline}
with a fluctuation current $I_{\rm F}(t) = 0$ and $\phi_{\rm J} + \phi = \phi_{\rm x}$. 
Here, $Z^{-1}(t)$ is the Fourier transform of a generalized admittance function $Z^{-1}(\omega)$ given by
\begin{align}\label{Gen_Admittance_eq}
Z^{-1}(\omega) = \sum_{n = 1}^{k}\frac{\omega_n}{i\omega L_n}\left \{ \frac{\omega_n}{(\omega+i\varepsilon)^2 - \omega_n^2} \right \}
=  \sum_{n = 1}^{k}\frac{\omega_n}{i\omega L_n}\left \{ \frac{1}{\omega - \omega_n + i\varepsilon} - \frac{1}{\omega + \omega_n + i\varepsilon} \right \}
\end{align}
\end{widetext}
\begin{align}\label{Sokhotski-Plemelj_eq}
\frac{1}{\omega + \omega_n \pm i\varepsilon} = \mp i\pi \delta(\omega + \omega_n) + {\rm p.p.}\left ( \frac{1}{\omega + \omega_n} \right ), 
\end{align}
\end{subequations}
where eq. (\ref{Sokhotski-Plemelj_eq}) is the Sokhotski-Plemelj formular and p.p. stands for Cauchy principal part. Eq. (\ref{Gen_Admittance_eq}) is related to the spectral function $K(\omega) = [Z^{-1}(\omega) + Z^{-1}(-\omega)]/(2\pi)$ given in section \ref{Environment} where $\varepsilon$ is the infinitesimal satisfying $\omega \varepsilon = \varepsilon$ and the nilpotent condition $\varepsilon^2 = 0$. Note, the spectral function is the sum of negative and positive frequency impedance accounting for emission and absorption processes respectively by the circuit. Thus, eq. (\ref{g_lagrangian_eq}) differs slightly from eq. (\ref{S_circuit_eq}) where the real-valued spectral function $K(t)$ in the classical Lagrangian gets replaced with the complex valued admittance $Z^{-1}(t)/(2\pi)$ in the quantum case. 

Introducing the Dirac delta function $\delta (x)$ for functional integrals with the property
\begin{equation}
\int Dx\, f(x) \delta(x-y)= f(y)
\label{delta_eq}
\end{equation}
for any functional $f(x)$, we may proceed to insert $\int D\phi_{\rm z}\delta(\phi_{\rm J}+\phi_{\rm x}+\phi_{\rm z}-e\Phi)=1$
into eq. (\ref{avrg_eq}) thus introducing the constraint $\sum_{a}\phi_{\rm a}=\phi_{\rm J}+\phi_{\rm x}+\phi_{\rm z}= e\Phi$ guaranteed by the circuit in Fig. (\ref{Caldeira_fig}). Consequently, the average in eq. (\ref{calculation_eq}) is now taken over both $\phi_{\rm J}$ and $\phi_{\rm z}$:  
\begin{widetext}
\begin{align}
\lim_{\beta \rightarrow +\infty}\langle \sin \left [ \kappa \Delta\phi_{\rm J}(t) \right ]\rangle_{\phi_{\rm J} \phi_{\rm z}}
= \mathcal{Z}^{-1}\int D\phi_{\rm J}\int D \phi_{\rm z}\delta\left ( \sum_{a}\phi_{\rm a}- e\Phi\right )
\sin \left [ \kappa \Delta\phi_{\rm J}(t) \right ]\exp i{S}'_{\rm z}(\phi - e\Phi).
\label{avrg2_eq}
\end{align}
We find,
\begin{multline}
-\langle \sin \left [ \kappa \Delta\phi_{\rm J}(t) \right ]\rangle_{\phi_{\rm J} \phi_{\rm z}} = \langle \sin [\kappa \Delta\phi_{\rm x}(t) + \kappa e\int_{0}^{t} A(\tau)d\tau + \kappa \Delta\phi_{\rm z}(t)]\rangle_{\phi_{\rm z}}\\
= \langle \sin \left [\kappa \Delta\phi_{\rm z}(t) \right ]\rangle_{\phi_{\rm z}}\cos \left [\kappa \Delta\phi_{\rm x}(t) + \kappa e\int_{0}^{t} A(\tau)d\tau \right]
\\ + \langle \cos \left [\kappa \Delta\phi_{\rm z}(t) \right ]\rangle_{\phi_{\rm z}}\sin \left [\kappa \Delta\phi_{\rm x}(t) + \kappa e\int_{0}^{t} A(\tau)d\tau \right],
\label{calculation2_eq}
\end{multline}
\end{widetext}
with $\Delta \Phi(t) = e\int_{0}^{t} A(\tau)d\tau$. We have assumed Fubini's theorem for interchange of integration order applies and thus performed first the integral over $\phi_{\rm z}$. Using the fact that $S'_{\rm z}$ is quadratic, the resulting functional integral over $\phi_{\rm z}$ in eq. (\ref{calculation2_eq}) is Gaussian resulting in $\langle \sin \left [\kappa \Delta\phi_{\rm z}(t) \right ]\rangle_{\phi_{\rm z}} = 0 $ term vanishing. Likewise, $\langle \cos \left [\kappa \Delta\phi_{\rm z}(t) \right ]\rangle_{\phi_{\rm z}}$ satisfies Wick's theorem\cite{Ingold_Nazarov1992} 
\begin{subequations}
\begin{align}\label{Wick_eq}
\langle \cos \left [\kappa \Delta\phi_{\rm z}(t) \right ]\rangle_{\phi_{\rm z}} = \exp\left ( {\kappa}^2\left \langle [\phi_{\rm z}(t)-\phi_{\rm z}(0)]\phi_{\rm z}(0) \right \rangle_{\phi_{\rm z}}  \right ),\\
\int D \phi_{\rm z}\exp iS'_{\rm z}(\phi_{\rm z}, I_{\rm F}) = \exp iS''_{\rm z}(I_{\rm F}),\\
{S}''_{\rm z}(I_{\rm F}) = \frac{2\pi}{2e^2}\int_{-\infty}^{+\infty}I_{\rm F}(-\omega)G_{\rm eff}(\omega)I_{\rm F}(\omega)d\omega\\
G_{\rm eff}(\omega) = -e^2 i\omega^{-1} Z_{\rm eff}(\omega),\\
Z_{\rm eff}(\omega) = \frac{1}{Z^{-1}(\omega) + y(\omega)}
\label{impedance_eq}
\end{align}
\end{subequations}
where $y(\omega) = i\omega C - i\omega^{-1}\sum_n L_n^{-1}$. 

We introduce the zero temperature propagator $D_{+\infty}(t)$ given by
\begin{subequations}\label{propagator_eq}
\begin{equation}
D_{+\infty}(t) = \frac{1}{2\pi }\int_{-\infty}^{+\infty}\frac{d\omega}{\omega}\exp-i\omega t
\left \{ Z_{\rm eff}(\omega) + n.f. \right \},
\end{equation}
where $n.f.$ stands for negative frequency. The finite temperature propagator is related to $D_{+\infty}(t)$ by a sum over the photon number states 
\begin{multline} 
D_{+\infty}(t) \rightarrow D_{\beta}(t) = \sum_{n = 0}^{+\infty}D_{+\infty}(t-in\beta) \\
= \frac{1}{2\pi}\int_{-\infty}^{+\infty}\frac{d\omega}{\omega}\frac{\exp-i\omega t}{1-\exp(-\beta \omega)}
\left \{ Z_{\rm eff}(\omega) + n.f. \right \} 
\label{propagator_finite_eq}
\end{multline}
\end{subequations}
Thus, computing the phase--phase correlation function, we find 
\begin{multline}\label{phase_phase_eq}
\left \langle \phi_{\rm z}(s)\phi_{\rm z}(t) \right \rangle_{\phi_{\rm z}}
= \left.\ \frac{\mathcal{Z}^{-1}\delta^2\exp iS''_{\rm z}(I_{\rm F})}{e^{-2}\delta I_{\rm F}(s)\delta I_{\rm F}(t)} 
 \right\vert_{{I_{\rm F}=0},\,{\mathcal{Z} = 1}}\\
= e^2D_{+\infty}(s-t) \rightarrow e^2D_{\beta}(s-t),
\end{multline}
which satisfies the well-know fluctuation-dissipation theorem.\cite{FDT1951} 

\subsubsection{\label{I_V} Cooper Pair and BCS quasi-particle Tunneling Current}

Finally, plugging in results (\ref{calculation2_eq}) and (\ref{propagator_finite_eq}) in eq. (\ref{calculation_eq}), and using $\Delta \phi_{\rm x}(t) = \int_{0}^{t}V(\tau)d\tau = Vt$ where $V_{\rm x} = V$ is a constant external voltage and $\Delta \Phi(t) = 0$, the total $I-V$ characteristics is given by 
\begin{widetext}
\begin{multline} \label{IV_eq}
I_{0}(V) = I_{1}(V) + I_{2}(V) \\
= e^{-1} R_{\rm T}^{-1}\int_{-\infty}^{+\infty}d\epsilon_{p}d\epsilon_{q}\frac{\mathcal{N}(\epsilon_{q})\mathcal{N}(\epsilon_{p})}{\mathcal{N}^2(0)}
f(\epsilon_{p})(1-f(\epsilon_{q}))\left \{P_1(\epsilon_{q}-\epsilon_{p} + eV) - P_1(\epsilon_{q}-\epsilon_{p} - eV) \right \}\\
+ e\pi E_{\rm J}^2\left \{ P_2(2eV)-P_2(-2eV) \right \} 
\end{multline}
where we have introduced the so called $P(E)$ function\cite{Ingold_Nazarov1992}
\begin{subequations} 
\begin{align}\label{P(E)_function_eq}
P_{\kappa}(E) =\frac{1}{2\pi} \int_{-\infty}^{+\infty}dt\, \exp\kappa^2 \mathcal{J}(t)\exp iEt,\\
e^{-2}\mathcal{J}(t) = D_{\beta}(t)-D_{\beta}(0)
\end{align}
\end{subequations}
with $E$ some arbitrary energy. It gives the probability that the junction will absorb energy $E$ from the environment.
Note that eq. (\ref{IV_eq}) reduces to the normal junction $I-V$ characteristics
\begin{multline}
I(V)|_{\Delta = 0} = e^{-1} R_{\rm T}^{-1}\int_{-\infty}^{+\infty}dE_{p}dE_{q}
f(E_{p})(1-f(E_{q}))
\left \{P_1(E_{q}-E_{p} + eV) - P_1(E_{q}-E_{p} - eV) \right \}\\
= e^{-1} R_{\rm T}^{-1}\int_{-\infty}^{+\infty} dE \frac{E}{1 - \exp(-\beta E)}
\left \{P_1(-E + eV) - P_1(-E - eV) \right \}
\label{normal_IV_eq}     
\end{multline}
when the superconducting gap vanishes $\Delta = 0$, since $E_{\rm J}(\Delta = 0) = 0$, $\mathcal{N}(\Delta = 0)/\mathcal{N}(0) = 1$ and $E_{p} = \epsilon_{p}, E_{q} = \epsilon_{q}$.


\section{\label{PP_Approach} The Array as an Effective Single Junction}

It is prudent to highlight the ingredients that went into deriving the $I$--$V$ characteristics of the single small Josephson junction given in eq. (\ref{IV_eq}): 

1) The Caldeira-Leggett action $S_{\rm z}(\phi'_{\rm x})$ that is varied with respect to $\phi = \phi_{\rm J} + \phi'_{\rm x}$ to obtain the equation of motion for the single $large$ Josephson junction; 2) the correlation $\langle \sin \left [ \kappa \Delta\phi_{\rm J}(t) \right ]\rangle_{\phi_{\rm J}}$ calculated with respect to $\phi_{\rm J}$; 3) The condition $\sum_i \phi_i = e\Phi$ enforced by the circuit. \\

In the case of a linear array of $N_0$ small Josephson junctions, it is clear that constraint 3) has to include all the phases $\phi_{{\rm J} = 1} \cdots \phi_{{\rm J} = N_{0}}$ of the junctions along the array, and the quantum average in 2) taken over each phase where the action in 1) is the sum of the action of individual junctions in the array. To simplify the calculation, one assumes all the junctions have the same structure coefficients $\alpha_{\kappa}(t)$ and calculates the quantum average
$\langle \sin \left [ \kappa \Delta\phi_{\rm J = 1} (t) \right ]\rangle_{\phi_{1} \cdots \phi_{N_0}}$ [step 2)]. Since the same current passes through all the junctions in the array, the calculation is carried out at any one of them [e.g. the ${\rm J} = 1$ junction] while assuming that the circuit forms a loop that imposes condition 3) as before. Hence, treating the other junctions as environments each with an effective action of the form $S'_{\rm z}(\phi_{\rm J})$, we have,
\begin{multline}
I_{\rm A}(V) = ie\sum_{\kappa = 1}^{2}  \int_{-\infty}^{+\infty}dt\,\alpha_{\kappa} (t)\langle \sin \left [ \kappa \Delta\phi_{\rm J = 1} (t) \right ]\rangle_{\phi_{\rm z}\phi_{1} \cdots \phi_{N_0}}\\
= -ei\sum_{\kappa = 1}^{2}\int_{-\infty}^{+\infty} dt\, \alpha_{\kappa}(t)\sin \left [ \kappa e\int_{0}^{t} A(\tau)d\tau + \kappa \Delta\phi_{\rm x} (t)\right]
\int \prod_{\rm z' = 1}^{N_{0}} D\phi_{\rm z'}\exp iS'_{\rm z}(\phi_{\rm z'})[\cos \kappa \Delta \phi_{\rm z'}(t)],
\label{I_V_array_eq}
\end{multline}
where,
\begin{multline*}
-\langle \sin \left [ \kappa \Delta\phi_{\rm J = 1} (t) \right ]\rangle_{\phi_{\rm z}\phi_{1} \cdots \phi_{N_0}}
= \langle \cos \left [ \kappa \Delta\phi_{\rm z} (t) \right ]\rangle_{\phi_{\rm z}}
\langle \sin \left [ \kappa \sum_{{\rm J} = 2}\Delta\phi_{\rm J} (t) +  \kappa e\int_{0}^{t} A(\tau)d\tau + \kappa \Delta\phi_{\rm x} (t)\right]   \rangle_{\phi_{2} \cdots \phi_{N_0}},\\
= \prod_{\rm z'}^{N_{0}}\langle \cos \left [ \kappa \Delta\phi_{\rm z'} (t) \right ]\rangle_{\phi_{\rm z'}}\sin \left [ \kappa e\int_{0}^{t} A(\tau)d\tau + \kappa \Delta\phi_{\rm x} (t)\right]
= \exp \left [-\frac{\kappa^2}{2}\sum_{\rm z'}^{N_0}\langle \Delta\phi_{\rm z'}^2 (t) \rangle_{\phi_{\rm z'}} \right ]\sin \left [ \kappa e\int_{0}^{t} A(\tau)d\tau + \kappa \Delta\phi_{\rm x} (t)\right],\\
= \exp \kappa^2\left [\sum_{\rm z'}^{N_0}\mathcal{J}_{\rm z'}(t)\right ]\sin \left [ \kappa e\int_{0}^{t} A(\tau)d\tau + \kappa \Delta\phi_{\rm x} (t)\right]
= \prod_{\rm z'}^{N_{0}} P_{\kappa}^{\rm z'}(t) \sin \left [ \kappa e\int_{0}^{t} A(\tau)d\tau + \kappa \Delta\phi_{\rm x} (t)\right].
\end{multline*}
Evidently, the current depends on the product
\begin{align*}
P_{\kappa}^{\rm A}(t) = \prod_{\rm z'}P_{\kappa}^{\rm z'}(t)   
\end{align*}
as expected, since, it is comprised of individual tunneling events at each junction. Since, $1/2\pi \int dt P_{\kappa}^{\rm A}(t)\exp iEt$ is the probability that the array will absorb energy $E$ from the environment, we discover that the tunnel current obeys the product rule of probabilities. For identical junctions of capacitance $C$ and impedance $Z(\omega)$,
\begin{align*}
\langle \phi_{\rm z}(t)\phi_{\rm z}(0)\rangle \equiv \langle \phi_{\rm z' = 2}(t)\phi_{\rm z' = 2}(0)\rangle
= \cdots = \langle \phi_{{\rm z'} = N_0}(t)\phi_{{\rm z'} = N_0}(s)\rangle,    
\end{align*}
we have $\prod_{\rm z'}^{N_0} P_{\kappa}^{\rm z'}(t) = [P_{\kappa}(t)]^{N_0}$ where
\begin{align}
[P_{\kappa}(t)]^{N_0} = 
\exp\left ( N_{0}{\kappa}^2\left \langle [\phi_{\rm z}(t)-\phi_{\rm z}(0)]\phi_{\rm z}(0) \right \rangle_{\phi_{\rm z}}  \right )
= \exp\left (N_{0}{\kappa}^2\mathcal{J}_{\rm A}(t) \right )
\label{P(E)_product_eq}
\end{align}
where $Z^{\rm A}_{\rm eff}(\omega)$ in $\mathcal{J}_{\rm A}(t) = \left \langle[\phi_{\rm z}(t)-\phi_{\rm z}(0)]\phi_{\rm z}(0)\right \rangle_{\phi_{\rm z}}$ will differ from $Z_{\rm eff}(\omega)$ in eq. (\ref{P(E)_function_eq}) due to possible interaction terms. Neglecting these interactions by setting $\mathcal{J}_{\rm A}(t) \simeq \mathcal{J}(t)$, eq. (\ref{P(E)_product_eq}) implies that at zero temperature, Cooper pair Coulomb blockade threshold voltage $V_{\rm cb}^{\rm A} = N_{0}V_{\rm cb}$ for the array is a factor $N_0$ larger than for the single junction.

However, the rest of the array ($\phi_{\rm J = 2}\cdots \phi_{{\rm J} = N_{0}}$) acts as the environment for the single junction ($\phi_{\rm J = 1}$) thus introducing interaction terms. In particular, the single junction interacts with the rest of the array electromagnetically. Since, in the presence of Cooper pair solitons\cite{Soliton_Schon1996, Delsing1992} and the Meisner effect, the electromagnetic field has a finite range within an infinitely long array ($N_{0} \gg 1$), the array has a cut-off number of junctions beyond which no electromagnetic interactions occur. Consequently, the effective number of junctions $N_{\rm c} \leq N_{0}-1$ acting as the environment will be determined by the range of the electromagnetic field. $N_{\rm c}(\Lambda)$ is independent of the magnetic field, $H$ when the superconducting islands are shorter than the penetration depth of the magnetic field, $H$. It can be evaluated by equating the Coulomb blockade voltage $V_{\rm cb}$ (estimated by replacing $N_{0}$ with $N_{\rm c}(\Lambda)$ and setting ${\rm Re} \left \{ Z_{\rm eff}^{\rm A}(\omega) \right \} \simeq {\rm Re}\left \{ Z_{\rm eff}(\omega) \right \}$ in eq. (\ref{P(E)_product_eq}) to the standard expression for the soliton threshold voltage\cite{Bakhvalov1989} of the array, $eV_{\rm cb}^{\rm A} = eV_{\rm th} \simeq 2E_{\rm c}[\exp(\Lambda^{-1})-1]^{-1}$, leading to \begin{subequations}\label{fict_bosons_eq}
\begin{align}
N_{0} \rightarrow N_{\rm c}(\Lambda) = \frac{1}{\exp(\Lambda^{-1}) - 1},
\end{align}
which approaches the soliton length $N_{\rm c}(\Lambda) \rightarrow \Lambda$ when $\Lambda \gg 1$. However, the infinite array effectively has
\begin{align}
N_{\rm c}(\Lambda) + 1 = \frac{1}{1 - \exp(-\Lambda^{-1})}
\label{cut_off_eq}
\end{align}
\end{subequations}
junctions. This means that $N_{0}$ in eq. (\ref{P(E)_product_eq}) is instead rescaled to $N_{\rm c}(\Lambda)$. Since $V_{\rm cb}^{\rm A}(\Lambda)$ should be invariant under the transformation $N_{\rm c}(\Lambda) \rightarrow N_{\rm c}(\Lambda) + 1$, we find
\begin{subequations}
\label{fict_bosons_trans_eq}
\begin{align}
\lim_{R \rightarrow +\infty} {\rm Re} \left \{ Z_{\rm eff}^{\rm A}(\omega) \right \} \simeq \lim_{R \rightarrow +\infty} {\rm Re}\left \{ Z_{\rm eff}(\omega) \right \}
\rightarrow \lim_{R \rightarrow +\infty} {\rm Re}\left \{ Z_{\rm eff}^{\rm A}(\omega) \right \}
\simeq \exp(-\Lambda^{-1})\,\lim_{R \rightarrow +\infty} {\rm Re}\left \{ Z_{\rm eff}(\omega) \right \} , 
\end{align}
\end{subequations}
we discover that switching on electromagnetic interactions leads to a rescaled impedance and a rescaled response function given by $\Xi(\omega) \rightarrow \Xi_{\rm A}(\omega)\,\Xi(\omega) = \exp(-\Lambda^{-1})\,\Xi(\omega)$. 

Finally, $N_{\rm c}$ [eq.  (\ref{cut_off_eq})] is given by [confer: eq.  (\ref{excitation_distribution_eq})],
\begin{equation}
N_{\rm c}(\Lambda) = \sum_{m = -\infty}^{+\infty} \frac{1}{\Lambda^{-1} - 2\pi m i } - \frac{1}{2} = \frac{1}{2}\coth\left ( \frac{1}{2\Lambda} \right ) - \frac{1}{2}. 
\label{N_c_eq}
\end{equation}
This result is not surprising, since we have determined the $I$--$V$ characteristics of the array by treating it as a single junction with the rest of the array acting as its environment. This means that the $N_0 - 1$ junctions themselves act as bosonic excitations whose (average) number $\langle N_0 \rangle - 1 = N_{\rm c}$ determines the electromagnetic cut-off, which is also the effective number of junctions that can be approximated as the environment of the effective single junction.

\section{\label{App:Functional_integral} Path Integral Formalism with Gaussian Functional Integral}

For completeness, this section summarizes how to compute correlation functions with Gaussian functional integrals such as the ones used in section \ref{path_integrals} in the derivation of the propagator $D_{+\infty}(t)$ in eq. (\ref{propagator_eq}). Our approach differs from typical procedures with imaginary time.\cite{Leggett1984} We work with real time instead since the finite temperature propagator is $trivially$ related to the zero temperature propagator [eq. (\ref{propagator_finite_eq})]. 

Consider a quadratic action $S(X, Y)$ with $X$ as the coordinate variable,  $Y$ as a fluctuation force, $a$ as a mass term and $g$ a coupling constant. The computation procedure is then as follows:
\begin{enumerate}
\item Take the Fourier transform of the action by substituting the Fourier or inverse Fourier transforms
\begin{align}
X(t) = \int d\omega X(\omega)\exp -i\omega t,\,X(\omega) = \frac{1}{2\pi}\int dt X(t)\exp i\omega t,
Y(t) = \int d\omega Y(\omega)\exp -i\omega t,\,Y(\omega) = \frac{1}{2\pi}\int dt Y(t)\exp -i\omega t
\end{align}
in the action,
\begin{align}
S(X, Y) = \int dt \left [ \frac{a}{2}\left ( \frac{\partial X(t)}{\partial t}\right )^2 - \frac{a}{2}\omega_0^2X^{2}(t) + gX(t)Y(t)\right ]
= 2\pi \int d\omega\left [\frac{1}{2}X(\omega)G^{-1}_{X}(\omega)X(-\omega) + gX(\omega)Y(-\omega)\right ],
\end{align}
where $a^{-1}G^{-1}_{X}(\omega) = (\omega + i\varepsilon)^2 - \omega_0^2$ and $G_{X}(t) = \int d\omega G_{X}(\omega) \exp(-i\omega t)$;
\item Perform the functional integral $\int DX \exp iS(X, Y) \propto \exp iS'(Y)$ emulating a typical Gaussian integral  
\begin{multline}
\int dx \exp i\left [a\frac{x^2}{2} + gyx\right ]
\propto \exp\left[i\frac{(ig)^2y^2}{2a}\right ] = \exp\left [-i\frac{g^2y^2}{2a}\right]\\ \rightarrow  S'(Y) = 2\pi \int d\omega \left [\frac{(ig)^2}{2}Y(\omega)G_{X}(\omega)Y(-\omega)\right ]
= \frac{(ig)^2}{2\times 2\pi}\int dtds\, Y(s)G_{X}(s-t)Y(t);
\end{multline}
\item Compute the correlation functions with the quadratic part of the action as follows,
\begin{multline}
\left \langle X(t_1) \cdots X(t_{n})\right \rangle 
= \mathcal{Z}^{-1}\int DX \left [ X(t_1) \cdots X(t_n) \right ]\exp iS(X,Y=0)\\
= \left [\frac{1}{(ig)^n}\frac{\delta}{\delta Y(t_n)} \cdots \frac{\delta}{\delta Y(t_1)}\mathcal{Z}^{-1}\int DX \exp iS(X,Y \neq 0)\right ]_{{Y = 0},\,{\mathcal{Z}=1}}\\
= \left [\frac{1}{(ig)^n}\frac{\delta}{\delta Y(t_n)} \cdots \frac{\delta}{\delta Y(t_1)}\exp iS'(Y)\right ]_{Y = 0}
= \left [\frac{1}{(ig)^n}\frac{\delta}{\delta Y(t_n)} \cdots \frac{\delta}{\delta Y(t_1)}\sum_{m = 0}^{m = \infty}\frac{\left \{ iS'(Y) \right \}^m}{m!}\right ]_{Y = 0}.
\end{multline}
We require the variation $\delta/\delta Y(t)$ to satisfy the anti-commutation rule,
\begin{subequations}
\begin{align}\label{var_rules_eq}
\frac{\delta}{\delta Y(t)}\frac{\delta}{\delta Y(s)}+ \frac{\delta}{\delta Y(s)}\frac{\delta}{\delta Y(t)}  = 0,
\end{align}
\end{subequations}
where $\delta Y(s)/\delta Y(t) = \delta(t - s)$ is the Dirac delta function. Note that the anti-commutation rule
accounts for time ordering. Since $S'(Y)$ is quadratic in $Y$, the integral vanishes for odd number of variables $n = 2N - 1$  where $N$ is a positive integer. For even number of variables $n = 2N$ we have the continuation,
\begin{multline}
\left \langle X(t_1) \cdots X(t_{n})\right \rangle = \left [\frac{i^N}{(ig)^nN!}\frac{\delta}{\delta Y(t_n)} \cdots \frac{\delta}{\delta Y(t_1)}S'^{N}(Y)\right ]
_{Y = 0,\,{n = 2N}}\\
= \frac{(ig)^{2N}}{(ig)^{n = 2N}}\frac{i^N}{N!}\frac{1}{(2\times 2\pi)^N}\frac{\delta}{\delta Y(t_n)} \cdots \frac{\delta}{\delta Y(t_1)}
\prod_{m = 1}^{m = N} \int ds_{2m - 1}ds_{2m}Y(s_{2m - 1})G_{X}(s_{2m - 1}-s_{2m})Y(s_{2m});
\end{multline}
\item For illustration, we compute the case $N = 1$,
\begin{multline}
\left \langle X(t_1)X(t_{2})\right \rangle_{t_1 \neq t_2} = \frac{i}{2\times 2\pi}\frac{\delta}{\delta Y(t_2)}\frac{\delta}{\delta Y(t_1)}\int ds_1 ds_2 Y(s_1)G_{X}(s_1-s_2)Y(s_2)\\
= \frac{i}{2\times 2\pi}\int ds_1 ds_2 \frac{\delta Y(s_1)}{\delta Y(t_2)}G_{X}(s_1-s_2)\frac{\delta Y(s_2)}{\delta Y(t_1)}
- \frac{i}{2\times 2\pi}\int ds_1 ds_2 \frac{\delta Y(s_1)}{\delta Y(t_1)}G_{X}(s_1-s_2)\frac{\delta Y(s_2)}{\delta Y(t_2)}\\
= \frac{i}{2\times 2\pi}\left \{ G_{X}(t_2 - t_1) - G_{X}(t_1 - t_2) \right \}
=\frac{i}{2\times 2\pi}\int d\omega \left [ G_{X}(\omega) - G_{X}(-\omega) \right ]\exp(-i\omega t)
\end{multline}
and
\begin{align}
\left \langle X(0)X(0)\right \rangle \equiv \left \langle X(t_1)X(t_2)\right \rangle_{t_1 = t_2} = \frac{i}{2\times 2\pi}\left \{ G_{X}(0^{+}) - G_{X}(0^{-}) \right \}
= \frac{i}{2\times 2\pi}\int d\omega \left [ G_{X}(\omega) - G_{X}(-\omega) \right ] \neq 0,
\end{align}
where $t = t_2 - t_1$ and we have used the anti-commutation rule given in 3. 
Note that, after Fourier transforming the action given in eq. (\ref{S_circuit_eq}) and the delta functional integral (\ref{delta_eq}) performed in eq. (\ref{calculation2_eq}), we simply have $X(t) \rightarrow \phi_{\rm z}(t)/\sqrt{2}$ and $G_{X}(\omega) \rightarrow G_{\rm eff}(\omega) = -e^2i\omega^{-1}Z_{\rm eff}(\omega)$ to yield eq. (\ref{phase_phase_eq}).
\end{enumerate} 

\end{widetext}

\end{appendix}

\end{document}